\documentclass[pre,twocolumn, superscriptaddress]{revtex4-1}

\usepackage{amsmath}
\usepackage{amsfonts}
\usepackage{amssymb}
\usepackage{array}
\usepackage{dcolumn}
\usepackage{longtable}
\usepackage{hyperref}

\usepackage{float}
\usepackage{graphicx}
\usepackage{natbib}
\usepackage{stmaryrd}%
\usepackage{bm}%
\usepackage{color}

\DeclareMathOperator*{\argmax}{arg\,max}%
\DeclareMathOperator*{\argmin}{arg\,min}%

\newcommand{\Up}{\mathbf{U}}
\newcommand{\Dn}{\mathbf{D}}

\renewcommand{\vec}[1]{\boldsymbol{#1}}

\DeclareRobustCommand{\binomial}{\genfrac(){0pt}{}}

\DeclareMathOperator\erf{erf}

\begin{document}

\title[Preisach model]%
{State transition graph of the Preisach model and the role of return point memory}

\date{\today}

\author{M. Mert Terzi}
\email{mert.terzi@u-psud.fr}
\affiliation{Universit\'e Paris-Saclay, CNRS, LPTMS, 91405 Orsay, France} 

\author{Muhittin Mungan}
\email{mungan@iam.uni-bonn.de}
\affiliation{Institut f\"{u}r angewandte Mathematik, Universit\"{a}t Bonn, Endenicher Allee 60, 53115 Bonn, Germany}

\begin{abstract}

The Preisach model has been useful as a null-model for understanding memory formation in periodically driven disordered systems. 
In amorphous solids for example, the athermal response to  shear is due to localized plastic events (soft spots). As shown recently by one of us, the plastic response to applied shear can be rigorously described in terms of a directed network whose transitions correspond to one or more soft spots changing states. The topology of this graph depends on the interactions between soft-spots and when such interactions are negligible, the resulting description becomes that of the Preisach model. 
A first step in linking transition graph topology with the underlying soft-spot interactions is therefore to determine the structure of such graphs in the absence of interactions. 
Here we perform a detailed analysis of the transition graph of the Preisach model. We highlight the important role played by return point memory in organizing the graph into a hierarchy of loops and sub-loops. Our analysis reveals that the topology of a large portion of this graph is actually not governed by the values of the switching fields that describe the hysteretic behavior of the individual elements,  but by a coarser parameter, a permutation $\vec{\rho}$ 
which prescribes the sequence in which the individual hysteretic elements change their states as the main hysteresis loop is traversed. 
This in turn allows us to derive combinatorial properties, such as the number of major loops in the transition graph as well as the number of states $\vert \mathcal{R} \vert$ constituting the main hysteresis loop and its nested sub-loops. We find that $\vert \mathcal{R} \vert$ is equal to the number of increasing subsequences contained in the permutation $\vec{\rho}$. 
\end{abstract}

%

\maketitle

\section{Introduction}
When  cyclically driven and under conditions where thermal effects are negligible, a wide variety of disordered condensed-matter systems  {\em anneal} by settling into a limit-cycle in which a set of microscopic configurations is visited periodically. 
Examples are magnets, dense colloidal suspensions, sheared amorphous solids and granular bead packs, which have been investigated extensively both experimentally as well as numerically \cite{Sethna93, Chaikin2008,Keimetal2011, Keimetal2013, Keimetal2014,regev2013onset, regev2015reversibility, Fioccoetal2014, Fiocco2015, Sastry2017, adhikari2018memory, KeimArratia2015, mukherji2018strength, Royer49, keim2018return, keim2018memory, KeimPaulsen2019, mungan2019networks}. 

In particular, in sheared amorphous solids these limit-cycles correspond to a repeating sequence of localized plastic events, referred to as shear transformation zones or soft-spots \cite{argon1979plastic,falk1998dynamics,manning2011softspot}. These soft-spots emerge as a result of the cyclic shearing and appear to be mostly  two-state systems with hysteresis. They 
interact with each other via long range quadrupolar displacement fields, typically associated with Eshelby inclusions \cite{ mungan2019networks,eshelby1957determination,maloney2006amorphous}.
As was shown by one of us recently \cite{mungan2019networks}, 
the primary effect of cyclic annealing is the formation of the interacting soft-spot system which in turn not only produces the periodic response, but at the same time renders it {\em resilient}: when the amplitude of the forcing is subsequently reduced the system often (but not always) settles into  a sub-cycle where a subset of the soft spots is active \cite{keim2018memory}. Indeed, experiments and numerical simulations have revealed that the interacting soft-spot system  formed by annealing gives rise to a hierarchy of cycles and sub-cycles that is moreover highly reminiscent of return point memory \cite{Barker1983,Sethna93,munganterzi2018, mungan2019networks, keim2018memory} (to be discussed further below). Such hierarchies were found to persist even at moderately large values of strain amplitudes \cite{mungan2019networks}.  

The states of the individual soft spots collectively encode the overall plastic configuration of the system via a {\em mesostate} \cite{mungan2019networks}. Mesostates are collections of configurations that under applied shear can be transformed into each other purely elastically. Plastic events then correspond to transitions between mesostates. As was shown in \cite{mungan2019networks}, it is possible to extract mesostates and their transitions from molecular statics simulations, such as those carried out in refs. 
\cite{regev2013onset,regev2015reversibility,Fioccoetal2014,Fiocco2015,Sastry2017, adhikari2018memory}. 
The description in terms of a state transition graph of mesostates allows one to relate features of the dynamic response, such as cycles and sub-cycles, to their corresponding graph theoretical counterparts \cite{mungan2019networks}. 

In the case of 
amorphous solids, the topology of the state transition graph is determined by how the soft spots interact with each other, namely how the state of a set of soft spots alters the switching behavior of another soft spot \cite{mungan2019networks}.   
However, a connection between graph topology and dynamics  is present
even when the soft-spots do not interact and hence switch independently of each other. 
This corresponds to the well-known Preisach model \cite{Preisach1935, mayergoyz1986mathematical, mayergoyz1986prl}. {This model} has been useful in understanding a broad range of systems exhibiting hysteresis, including magnetic materials, where the model originated \cite{Preisach1935, everett1954general, Barker1983}, but also fracture in dilatant rocks \cite{holcomb1981memory, claytor2009limitations, scalerandi2010nonequilibrium}, and more generally, memory formation in matter \cite{schubert2017preisach,keim2018memory, keim2018return, KeimPaulsen2019}. A comprehensive review of the Preisach model and its applications can be found in  \cite{bertotti2006book} and {\cite{brokate2012hysteresis}}. 

The goal of this paper is therefore twofold. On the one hand,  treating the Preisach model as a null-model of non-interacting soft-spots forming the limit-cycle of a periodically sheared amorphous solid, we aim to describe its transition graph.  {While transition graphs for some particular and special realizations of the Preisach model have been considered before \cite{rachinskii2016realization,kalmar2019devil}, our goal here is to provide a general description of all possible transition graphs}. Comparing the structure of such graphs with those extracted from limit-cycles of real systems (numerically or by experiment), any deviations from the topology of the Preisach graphs can be attributed to interactions between the soft-spots. This will be useful in identifying non-trivial network motifs in state transition graphs that have been extracted from systems with interactions, as was done in \cite{mungan2019networks}.  

On the other hand, the Preisach model is the simplest system exhibiting return point memory (RPM) \cite{everett1954general, Barker1983, Sethna93}, a property wherein a system remembers the states at which the direction of an external driving had been reversed. As we have shown recently \cite{munganterzi2018}, the presence of RPM imposes strong constraints on the topology of the associated state transition graph. Thus many features of the state transition graph associated with the Preisach model, and hence its dynamics, are a direct consequence of RPM. Our second goal therefore is to use the Preisach model to illustrate some of the theoretical results 
that we obtained before \cite{munganterzi2018}.  
%

We conclude this introduction with a summary of our main results, {(P1) - (P4)},  and an outline of the paper. 
In Section \ref{sec:defs} we define the Preisach model, its stable states and the transitions between them. Given a stable configuration, there exists a minimal field change so that a new stable state is always reached by the state change of a single hysteretic element (hysteron). In other words, the Preisach model does not exhibit avalanches. We refer to the set of field values at which the individual hysterons switch their states as the switching fields.  In subsection \ref{subsec:rhostable} we introduce the idea of a 
$\rho$-stable state. A $\rho$-stable state is a hysteron configuration that is stable for any realization of the switching fields  whose ordering is compatible with a given permutation $\vec{\rho}$. In particular, we show that all {\em 2-loops}, {\em i.e. } pairs of states that transition into each other under  the state change of a single hysteron, are $\rho$-stable -- property (P1). In Section \ref{sec:rpm} we summarize the general theory of state transition graphs obeying the return point memory property, as worked out by us in \cite{munganterzi2018}, introducing the {\em loop-RPM} property ($\ell$RPM) and  maximal loops. We then turn to an application of these results to the Preisach model. 
In section \ref{sec:preisachmax} we describe the maximal loops of the Preisach model and generalize (P1) by  establishing $\rho$-stability for all states associated with non-trivial maximal loops, {\em i.e.} loops having at least two states -- properties (P2) and (P3). {As a consequence, the states associated with the main hysteresis loop of the Preisach model and all its sub-loops are $\rho$-stable}. In Section \ref{sec:preigraph} {we describe the transition graph of the main hysteresis loop,  the {\em Preisach graph}}. We show that the 
permutation $\vec{\rho}$ completely determines its topology, and conversely, that given the Preisach graph, $\vec{\rho}$ can be read off from its topology. Having established the combinatorial structure underlying the transitions graph of the Preisach model, we turn next in Section \ref{sec:statistics} to the derivation of two statistical results: (i) the disorder-averaged size distribution of maximal loops, and (ii) the number of reachable states of the main hysteresis loop, {\em i.e.} the number of configurations reachable from one of the saturated states. Given the permutation $\vec{\rho}$, we show that this number is equal to the set of increasing subsequences contained in $\vec{\rho}$ -- property (P4). Using this observation, we establish that the disorder-averaged number of reachable states is asymptotic to $L^{-1/4} \exp{(2 L^{1/2})}$, where $L$ is the number of hysterons of the model. 
We conclude with a discussion of our results in Section \ref{sec:discussion}. Appendix \ref{app:2loop} contains the  proof of $\rho$-stability of two-loops, while in Appendix \ref{sec:NP} we derive the no-passing (NP) property for the Preisach model \cite{NoPassing}, which implies the RPM property  \cite{Sethna93}. 

\section{The Preisach model as an AQS automaton}
\label{sec:defs}

\subsection{Definitions}

We start with the definition of the Preisach model \cite{Preisach1935}. We consider a collection of $L$ two-level systems, the {\em hysterons}. 
The configurations are $L$-component 
vectors $\vec{\sigma}$, with components $\sigma_i = \pm 1$, designating the state of each hysteron. With every hysteron $i$ we associate 
a pair of real numbers $(F^-_i, F^+_i)$, the {\em switching fields}, satisfying
\begin{equation}
 F^-_i  < F^+_i.
 \label{eqn:tipF}
\end{equation}
If hysteron $i$ is in state $\sigma_i = +1$, then it will become unstable when $F \leq F^-_i$ and switch to the state $\sigma_i = -1$. 
Likewise, if  $i$ is in state $\sigma_i = -1$, it will become unstable and switch to $\sigma_i = +1$ when  
$F \geq F^+_i$. Thus a hysteron $i$ is stable at force $F$ if either $\sigma_i = -1$ and $F < F^+_i$ or $\sigma_i = 1$ and $F > F^-_i$.
We assume that the switching fields $F^\pm_i$ are all 
distinct. {Note that this assumption is natural in the context that we are interested in, {\em e.g.} sheared amorphous solids \cite{mungan2019networks,argon1979plastic,falk1998dynamics,manning2011softspot}. Here one considers a {\em finite}, but possibly large, number of hysteretic elements, whose switching properties are due to some underlying microscopic disorder. Thus the switching fields can be assumed to be distinct \footnote{The point of view underlying the standard treatment of Preisach models is more macroscopic: One considers a continuum of hysteretic elements, so that a realization of the Preisach model is described by a continuous {\em density} of switching fields. Likewise, in such a setting the output variable of interest is not the state of each individual hysteron anymore, but a macroscopic quantity such as the average magnetization, in the case the hysterons carry a magnetic moment. Details of such treatments can be found for example in \cite{bertotti2006book} and \cite{brokate2012hysteresis}.}. }         

Given a configuration $\vec{\sigma}$, let the sets $I^\pm[\vec{\sigma}]$ denote the collections of  hysteron $i$ that are in state $\pm 1$:
\begin{align}
 I^+[\vec{\sigma}] &= \{ i : \sigma_i = +1\}, \label{eqn:iplusdef} \\
 I^-[\vec{\sigma}] &= \{ i : \sigma_i = -1\}. \label{eqn:iminusdef}
\end{align}
We define next the threshold fields $F^\pm[\vec{\sigma}]$ associated with a configuration $\vec{\sigma}$ as: 
\begin{align}
  F^+[\vec{\sigma}] &= \min_{i \in \mathcal{I}^-[\vec{\sigma}]} F^+_i, \label{eqn:trapDefp} \\
  F^-[\vec{\sigma}] &= \max_{i \in \mathcal{I}^+[\vec{\sigma}]} F^-_i. \label{eqn:trapDefm}
\end{align}
Let $\vec{\alpha}$ and $\vec{\omega}$ be the states $(-1,-1, \ldots, -1)$ and $(+1,+1, \ldots, +1)$, respectively. In magnetism language 
these are the saturated states. It is convenient to set
\begin{equation}
F^+[\vec{\omega}] = \infty, \quad \quad F^-[\vec{\alpha}] = -\infty.
\label{eqn:gamma_ep}
\end{equation}

In the Preisach model, a configuration $\vec{\sigma} = (\sigma_1, \sigma_2, \ldots, \sigma_L)$ is stable, if there is 
a field $F$ at which each hysteron $i$ is stable, in the sense defined above. It is easily shown that this condition is equivalent to 
requiring that 
\begin{equation}
 F^-[\vec{\sigma}] < F^+[\vec{\sigma}]. 
 \label{eqn:defStable}
\end{equation}
We denote the set of all stable states by $\mathcal{S}$ and write its number of elements as $\vert \mathcal{S} \vert$. In the AQS (athermal quasi-static) regime we are interested only in the stable states and the transitions between them. Note that the states $\vec{\alpha}$ and $\vec{\omega}$ are always stable, {\em i.e.} irrespective of the choice of switching fields $F^\pm_i$. 

Given a stable state $\vec{\sigma}$, we denote by $i^+[\vec{\sigma}]$ or $i^-[\vec{\sigma}]$ as, {\em cf.} \eqref{eqn:trapDefp} and \eqref{eqn:trapDefm}, 
\begin{align}
    i^+[\vec{\sigma}] &= \argmin_{i \in \mathcal{I}^-[\vec{\sigma}]} F^+_i, \label{eqn:trapDefip} \\
  i^-[\vec{\sigma}] &= \argmax_{i \in \mathcal{I}^+[\vec{\sigma}]} F^-_i. \label{eqn:trapDefim}
\end{align}
Since the switching fields $F^\pm_i$ of the individual hysterons were assumed to be  distinct, the sites $i^\pm[\vec{\sigma}]$ are unique \footnote{Note that when $\vec{\sigma} = \vec{\omega}$  $(\vec{\sigma} =  \vec{\alpha})$, the set $I^-$ ($I^+$), is empty and so $i^+$ ($i^-$) is undefined. In these cases no transitions occur, so this is consistent.}. 
They are the 
least stable hysterons with respect to field decrease and increase, respectively and will change state when $F = F^\pm[\vec{\sigma}]$. 

Given a stable state $\vec{\sigma}$, and an initial force $F$ such that 
$F^-[\vec{\sigma}] < F < F^+[\vec{\sigma}]$, we are interested in the transition into another stable state when the field is raised to $F^+[\vec{\sigma}]$ (and kept constant). At least one hysteron, namely $i^+$ will change its state in this case. We show next that the state change of this one hysteron suffices to obtain a new state $\vec{\sigma'}$ that is stable at the force $F = F^+[\vec{\sigma}]$. To see this, let $\vec{\sigma}$ be stable and assume that $\vec{\sigma} \ne \vec{\omega}$. The set $I^-[\vec{\sigma}]$ is thus non-empty and there is a least stable site $i^+[\vec{\sigma}]$ that will become unstable when $F = F^+[\vec{\sigma}]$, since by definition $F^+[\vec{\sigma}] = F^+_{i^+[\vec{\sigma}]}$. 
Let $\vec{\sigma'}$ denote the configuration obtained from $\vec{\sigma}$ by changing only the state of 
hysteron $i^+$ from $-1$ to $+1$. We claim that $\vec{\sigma'}$ is stable. Observe that $I^+[\vec{\sigma'}] = I^+[\vec{\sigma}] \cup \{ i^+[\vec{\sigma}] \}$. From \eqref{eqn:trapDefm} it immediately follows that 
\begin{equation}
    F^-[\vec{\sigma'}] \le F^-[\vec{\sigma}]. 
\end{equation}
Likewise, we have $I^-[\vec{\sigma'}] = I^-[\vec{\sigma}] \setminus \{ i^+[\vec{\sigma}] \}$, so that 
\begin{equation}
    F^+[\vec{\sigma'}] \ge F^+[\vec{\sigma}]. 
\end{equation}
Since $\vec{\sigma}$ is stable by assumption and \eqref{eqn:defStable} holds, it follows that $F^-[\vec{\sigma'}] < F^+[\vec{\sigma'}]$ and hence 
$\vec{\sigma'}$ is stable, too. The case for a transition when $F$ is lowered to $F^-[\vec{\sigma}]$ proceeds similarly. 
We have thus re-derived the following property of the Preisach model: 
\begin{enumerate}
    \item [{(P)}] A stable state $\vec{\sigma} \in \mathcal{S}$  transits at $F = F^-[\vec{\sigma}]$  ($F = F^+[\vec{\sigma}])$ via the state change of a {\em single} hysteron $i^-$ ($i^+$) to a new stable state. 
\end{enumerate}
Property (P) implies that the Preisach model does not have avalanches: as the forcing is lowered or raised, hysterons change their state one at a time. 

\subsection{The maps $\Up, \Dn$ and the state transition graph}
Given a stable state $\vec{\sigma}$ and setting $F = F^+[\vec{\sigma}]$, allows us to define a map {that} takes a stable state $\vec{\sigma}$ to another stable state $\vec{\sigma'}$ under minimal field increase. We write this as 
\begin{equation}
\vec{\sigma'} = \Up \vec{\sigma}. 
\end{equation}
Similarly, we define the transition from a state $\vec{\sigma}$ under a field decrease to $F = F^-[\vec{\sigma}]$ in terms of a map $\Dn$ as
\begin{equation}
\vec{\sigma''} = \Dn \vec{\sigma}. 
\end{equation}

With \eqref{eqn:gamma_ep}, it is convenient to define the $\Up$- ($\Dn$)-transitions from $\vec{\omega}$ ($\vec{\alpha}$) as 
\begin{equation}
 \Up \vec{\omega} = \vec{\omega}, \quad \quad \Dn \vec{\alpha} = \vec{\alpha}. 
 \label{eqn:aoabsorption}
\end{equation}
In this way both $\Up$ and $\Dn$ map $\mathcal{S}$ into itself. 

The maps $\Up$ and $\Dn$ together with the pair of switching  fields $F^\pm[\vec{\sigma}]$ of each stable configuration suffice to determine the AQS response of the Preisach model to arbitrary force protocols $F(t)$ \cite{munganterzi2018,munganwitten2018}. We refer to such systems as {\em AQS-automata}. To see this, suppose the system is initially in a state $\vec{\sigma}$ that is stable at $F_0$. When we  increase the force  to some value $F${, t}he system will transit through the states $\vec{\sigma}, \Up \vec{\sigma}, \Up^2 \vec{\sigma}, \ldots$, until reaching the first state $\Up^k \vec{\sigma}$, for which 
$F^+[\Up^k \vec{\sigma}] > F$. 

\begin{figure}[t!]
  \begin{center}
    \includegraphics[width=\columnwidth]{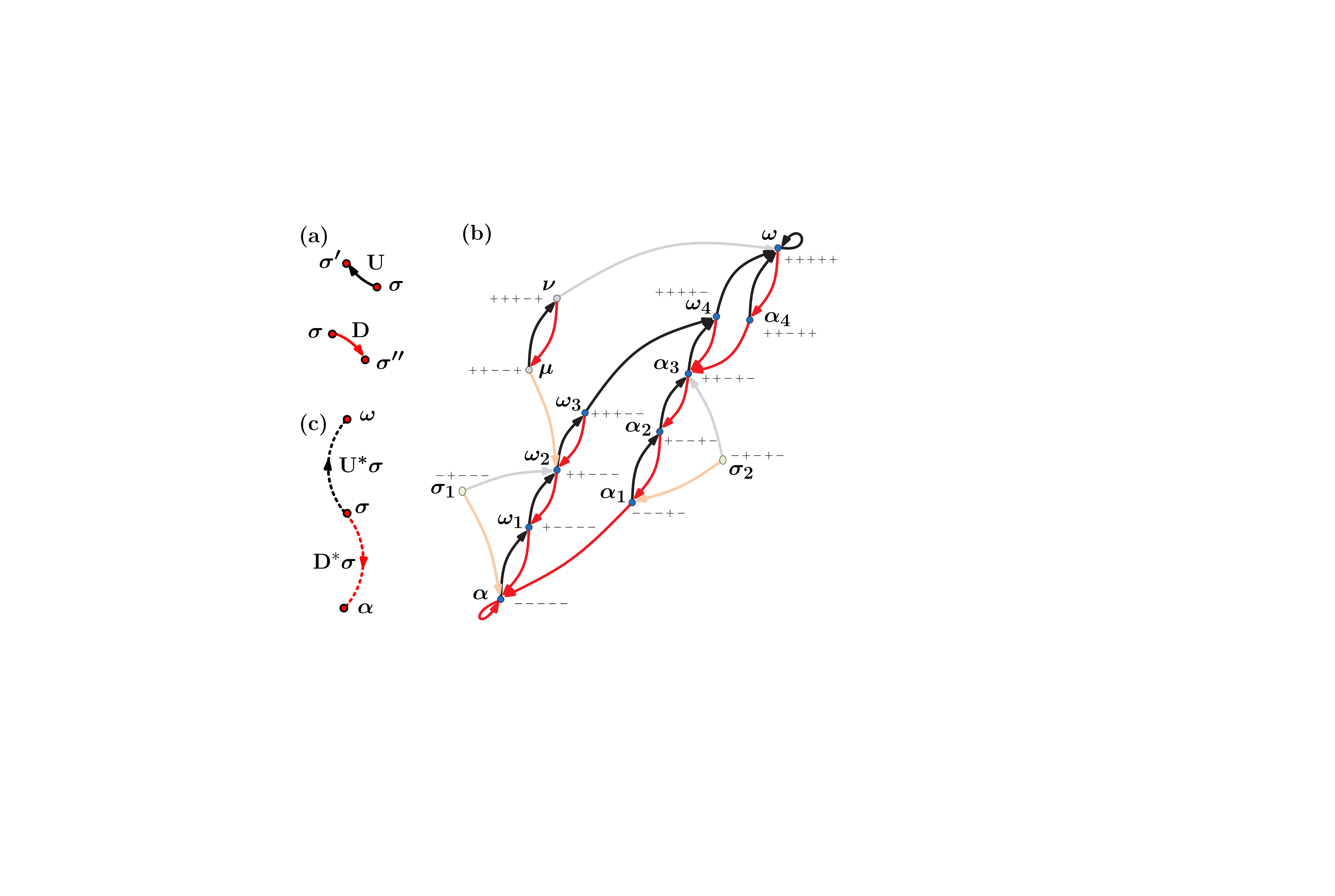} 
  \end{center}
  \caption{(color online) (a) The graph representation of the the actions of $\Up$ and $\Dn$, upward pointing arrows (black) and downwards pointing (red) arrows, respectively. This convention for pointing direction (color) will be used in all subsequent figures, substituting at times gray shades (pale colors) to de-emphasize transitions.  (b) 
  The state transition graph of the Preisach model with $5$ hysterons and ordering permutation $\vec{\rho} = (35214)$. The vertices associated with the main hysteresis loop and its sub-loops are shown in blue, and are labeled using the letters  $\vec{\alpha}, \vec{\alpha_1}, \vec{\alpha}_2, \vec{\alpha}_3, \vec{\alpha}_4$, and $\vec{\omega}, \vec{\omega_1}, \vec{\omega}_2, \vec{\omega}_3, \vec{\omega}_4 $. They constitute the set of reachable states $\mathcal{R}$, 
  {\em i.e.} the states reachable from one of the two saturated configurations $\vec{\alpha}$ and $\vec{\omega}$ by a sequence of $\Up$ and $\Dn$ transitions. 
  {The states $\vec{\mu}, \vec{\nu}, \vec{\sigma}_1$, and $\vec{\sigma}_2$ are not reachable.} The pair of states 
  $(\vec{\mu}, \vec{\nu})$ form a maximal $2$-loop, as does the the main hysteresis loop  $(\vec{\alpha}, \vec{\omega})$,   and the two singleton maximal loops consisting of the states  $\vec{\sigma_1}$ and $\vec{\sigma_2}$, respectively. It turns out that the ordering permutation $\vec{\rho}$ suffices to describe the set of all states that are non-singleton maximal loops as well as the transitions between them (refer to text for details).   
  (c) The $U$- and $D$-orbits of a state $\vec{\sigma}$ will be designated by dashed arrows {(refer to Section \ref{sec:rpm} for details)}.} 
  \label{fig:Gen_defs}
\end{figure}

The maps $\Up$ and $\Dn$ each have a natural representation as a directed graph, called the functional graph of the map. Its vertex set is $\mathcal{S}$ and its directed edges are the sets of pairs $(\vec{\sigma},\Up \vec{\sigma})$ and $(\vec{\sigma}, \Dn \vec{\sigma})$, respectively. These transitions can be represented as directed arrows connecting the initial and final states, as shown in Fig.~\ref{fig:Gen_defs}(a). We shall adopt the convention to mark transitions under $\Up$ and $\Dn$ by {upward pointing black and downward pointing red arrows}, respectively, using the lighter colors gray and orange if we want to emphasize these less. Combining both sets of edges, we obtain the multi-graph on $\mathcal{S}$, the AQS state transitions graph. It governs the dynamics of the Preisach model under arbitrary forcing, as outlined above. Fig.~\ref{fig:Gen_defs}(b) shows the state transition graph for a Preisach system with $5$ hysterons, giving rise to 14 stable states. 

\subsection{The ordering permutation $\rho$ and $\rho$-stability}
\label{subsec:rhostable}

In order to simplify notation for the rest of the paper we assume without loss of generality that the hysterons $i = 1, 2, \ldots, L$ have been indexed such that 
\begin{equation}
 F^+_1 < F^+_2 < \ldots < F^+_L.
 \label{eqn:Fplus_order}
\end{equation}
Denote by 
\begin{equation}
    \vec{\rho} = \left ( 
    \begin{array}{cccc}
    1 & 2 & \cdots & L \\ \rho_1 & \rho_2 & \ldots & \rho_L 
    \end{array}
    \right )
    \label{eqn:permdef}
\end{equation}
the permutation  ordering the $F^-_i$ from largest to smallest:
\begin{equation}
 F^-_{\rho_1} > F^-_{\rho_2} > \ldots > F^-_{\rho_L}. 
  \label{eqn:rhoorder}
\end{equation}
In a slight abuse of notation we will write the permutation in  \eqref{eqn:permdef} as $\vec{\rho} = (\rho_1, \rho_2, \ldots, \rho_L)$.  

As remarked in the introduction, one of our main findings is that the topology of a structurally large portion of the transition graph for the Preisach model  
depends only on the relative order of the switching fields, as specified by the permutation $\vec{\rho}$ and not their specific values $F^\pm_i$ -- as long as the inequality \eqref{eqn:tipF} holds, of course. This means that there is a subset of hysteron configurations whose stability depends entirely on $\vec{\rho}$ and hence remain stable for any realization of switching fields $F^\pm_i$ compatible with $\vec{\rho}$. Let us call  such states {\em $\rho$-stable} and denote their set by $\mathcal{S}_{\rho}$. 

The existence of $\rho$-stable states might seem counter-intuitive, since one expects that whether a hysteron configuration $\vec{\sigma}$ is stable or not, and hence satisfies \eqref{eqn:defStable}, should depend via \eqref{eqn:trapDefp} and \eqref{eqn:trapDefm} on the particular values of the switching fields. 
For example, if the $F^\pm_i$ were to satisfy the stronger condition 
\begin{equation}
    F^-_i < F^+_j
    \label{eqn:strongCondition}
\end{equation}
for all $i,j$ (which can be realized by requiring that $F^-_i < 0 < F^+_j$), then all $2^L$ hysteron configurations are stable, as is readily shown. It is well-known however that for general choices of $F^\pm_i$ satisfying \eqref{eqn:tipF}, not all $2^L$ possible hysteron configurations are stable. 

We will call the ordered pair of stable states $(\vec{\sigma}, \vec{\sigma'})$ a {\em two-loop}, if the following hold
\begin{align}
    \vec{\sigma'} &= \Up \vec{\sigma}, \\
    \vec{\sigma} &= \Dn \vec{\sigma'}.
\end{align}
{For example, $(\vec{\mu},\vec{\nu})$ and $(\vec{\alpha}_3,\vec{\omega}_4)$ in Fig.~\ref{fig:Gen_defs}(b) are two-loops.}
Denote by $\mathcal{S}^{(2)}$ the set of all states that are part of some two-loop. We derive next the following property of the Preisach model:

\begin{itemize}
    \item [(P1)] All two-loops are $\rho$-stable,
    \begin{equation}
        \mathcal{S}^{(2)} \subset \mathcal{S}_{\rho}.
        \label{eqn:S2subsetSrho}
    \end{equation}
\end{itemize}

Observe that for any state $\vec{\sigma} \in \mathcal{S}^{(2)}$, we must either have that 
\begin{equation}
 \vec{\sigma} = \Dn \Up \vec{\sigma}, 
 \label{eqn:DUsigma}
\end{equation}
or
\begin{equation}
 \vec{\sigma}  = \Up \Dn \vec{\sigma}, 
 \label{eqn:UDsigma}
\end{equation}
If $\vec{\sigma}$ satisfies \eqref{eqn:DUsigma}, then $\vec{\sigma'} = \Up \vec{\sigma}$ satisfies \eqref{eqn:UDsigma}. Moreover, by property (P) established earlier, if $\vec{\sigma}$ is stable, then $\Up \vec{\sigma}$ has to be stable as well. Thus it suffices to consider only the set of states $\vec{\sigma}$ satisfying \eqref{eqn:DUsigma} and show that these states belong to $\mathcal{S}_{\rho}$. It is readily shown, {\em see} Appendix \ref{app:2loop}, that any state $\vec{\sigma}$ satisfying \eqref{eqn:DUsigma} must be of the form  
\begin{equation}
    \sigma_i = \left \{ \begin{array}{cc}
    +1, & i < k, \\
    -1, & i \in \{ \rho_1, \rho_2, \ldots, \rho_r \}, \\
    *, & \mbox{otherwise},
    \end{array}
    \right.
    \label{eqn:2loop2}
\end{equation}
with $k = \rho_r$ chosen such that it is lower record of $\vec{\rho}$,  
\begin{equation}
     \rho_r = \min_{1 \le u \le r} \rho_u.
\end{equation}
Moreover, such a state will always be stable {and property (P1) has been proven}.

\section{State transition graphs of AQS systems with return point memory}  
\label{sec:rpm}

In the previous section we have cast the dynamics of the Preisach model in terms of a set of configuration $\mathcal{S}$ and two maps $\Up$ and $\Dn$ that map $\mathcal{S}$ into itself. Such a description emerges naturally in the AQS regime \cite{maloney2006amorphous} in which one considers the athermal and adiabatic response of a driven disordered system  \cite{munganterzi2018, munganwitten2018}. When 
such systems exhibit the return point memory (RPM) property, the corresponding state transition graph possesses a certain topological structure called $\ell$RPM, {which we} have identified in \cite{munganterzi2018}. 

The "classical" route to RPM is via Middleton's no-passing (NP) property \cite{NoPassing,Sethna93}. In the context of AQS dynamics, a system possesses the NP property, if there exists a partial order $\preceq$ on the set of stable configurations $\mathcal{S}$ which is preserved by the dynamics. Details have been given in Appendix \ref{sec:NP}, where we also provide a proof that the AQS dynamics of the Preisach model exhibits the NP property. Consequently, the state transition graph of the Preisach model exhibits $\ell$RPM. Note that the reverse is not true in general, an AQS system whose state transition graph exhibits the $\ell$RPM topology, does not necessarily have to satisfy an NP property. 

In the following we review the structure of state transition graphs with the $\ell$RPM property. Details and proofs of the statements {presented} here can be found in \cite{munganterzi2018}.  

Let $\vec{\sigma}$ be any stable state. The sequence of states obtained by repeatedly applying $\Up$ is called the $\Up$-orbit of $\vec{\sigma}$ and we write this as  \footnote{We are assuming here that the AQS system has unique absorbing states $\vec{\alpha}$ and $\vec{\omega}$, under $\Dn$ and $\Up$, respectively, as is the case for the Preisach model.} 
\begin{equation}
  \Up^* \vec{\sigma} =    (\vec{\sigma}, \Up \vec{\sigma}, \Up^2 \vec{\sigma}, \ldots , \vec{\omega}).
\end{equation}  
Likewise, the $\Dn$-orbit of $\vec{\sigma}$ is defined as 
\begin{equation}
  \Dn^* \vec{\sigma} =    (\vec{\sigma}, \Dn \vec{\sigma}, \Dn^2 \vec{\sigma}, \ldots , \vec{\alpha}).
\end{equation}
Fig.~\ref{fig:Gen_defs}(c) depicts the graphical representation we shall use for orbits. 

A pair of states $(\vec{\mu}, \vec{\nu})$ forms a {\em  loop}, if $\vec{\nu} \in \Up^* \vec{\mu}$ and 
simultaneously $\vec{\mu} \in \Dn^* \vec{\nu}$. We call $\vec{\mu}$ and $\vec{\nu}$ the lower and upper endpoint of the loop, respectively. Moreover, there are (smallest) integers $n,m \ge 0$ such that 
$\vec{\nu} = \Up^n \vec{\mu}$ and $\vec{\mu} = \Dn^m \vec{\nu}$. 
The states $\vec{\nu}_i = \Up^i \vec{\mu}$ with $i = 0, 1, 2, \ldots, n$, and $\vec{\mu}_j = \Dn^{m - j} \vec{\nu}$ for $j = 0, 1, 2, \ldots, m$, are called the $\Up$-, respectively $\Dn$-boundary states of {the loop $(\vec{\mu}, \vec{\nu})$. For example, in Fig.~\ref{fig:Gen_defs}(b), 
the pair of states $(\vec{\alpha}, \vec{\omega})$ forms a loop and its $\Up-$, respectively $\Dn$-boundary states are the set of states 
$\{\vec{\alpha}, \vec{\omega}_1, \vec{\omega}_2, \vec{\omega}_3, \vec{\omega}_4, \vec{\omega}\}$ and $\{\vec{\alpha}, \vec{\alpha}_1, \vec{\alpha}_2, \vec{\alpha}_3,  \vec{\alpha}_4, \vec{\omega}\}$, respectively. }

\begin{figure}[t!]
  \begin{center}
    \includegraphics[width=\columnwidth]{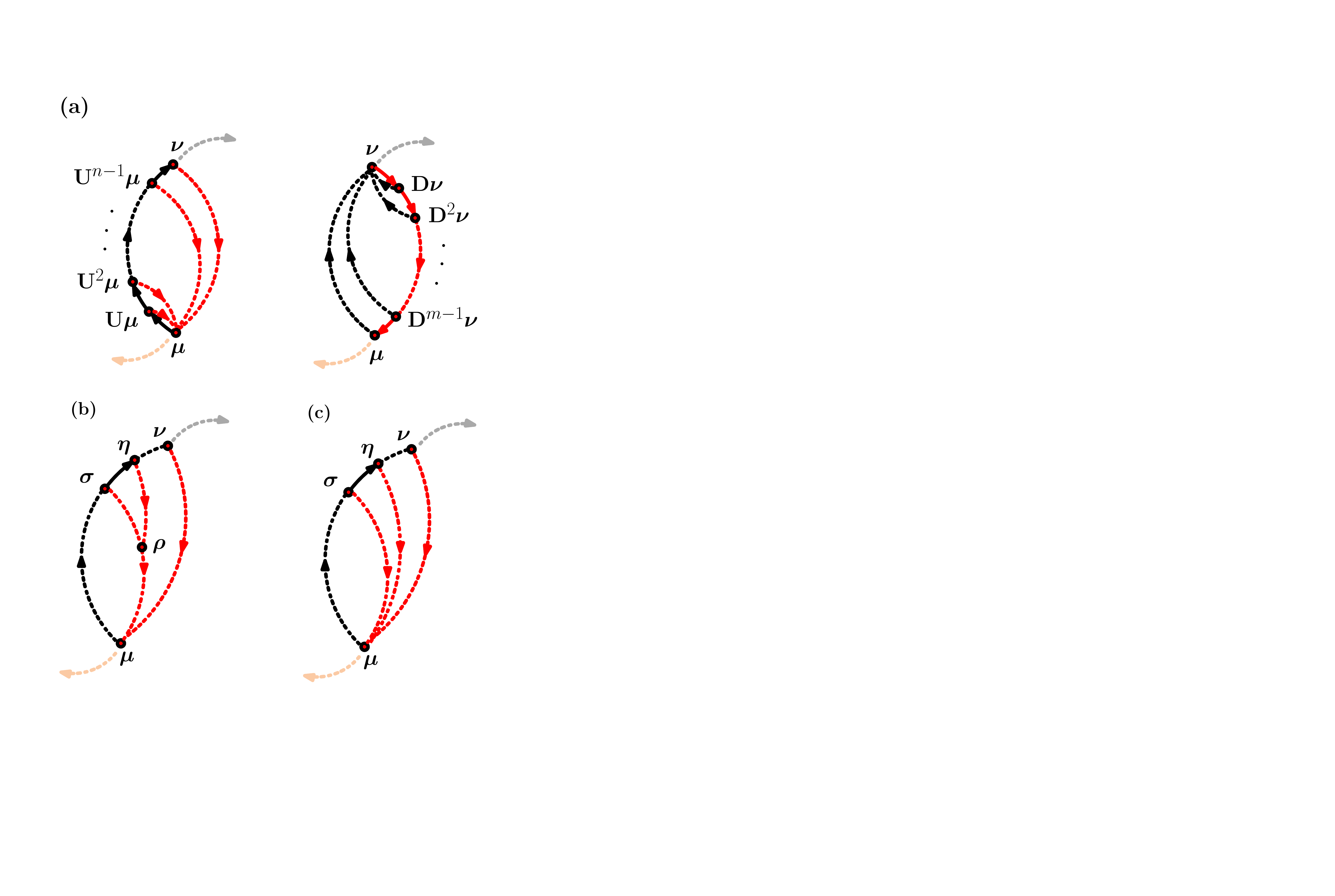} 
  \end{center}
  \caption{(color online)(a) Illustration of the absorption property of a loop $(\vec{\mu}, \vec{\nu})$. The $\Up$- and $\Dn$-boundary states of the loop are labeled as 
  $\Up^i \vec{\mu}$ and $\Dn^j \vec{\nu}$, respectively. The loop $(\vec{\mu}, \vec{\nu})$ is {\em absorbing} if all $\Dn$-orbits off the 
  $\Up$-boundary pass through the lower endpoint $\vec{\mu}$ of the loop (left panel), and likewise all $\Up$-orbits off the $\Dn$-boundary pass through $\vec{\nu}$ (right panel). The nesting property of orbits is illustrated in (b) and (c). The states $\vec{\sigma}$ and $\vec{\eta} = \Up \vec{\sigma}$ are two successive $\Up$-boundary states {of the loop $(\vec{\mu}, \vec{\nu})$}. By the absorption property {of $(\vec{\mu}, \vec{\nu})$}, 
  their $\Dn$-orbits must lead to $\vec{\mu}$. From the $\ell$RPM property it follows that these two orbits must merge either 
  prior to reaching $\vec{\mu}$ at some state $\vec{\rho}$, as shown in (b), or at $\vec{\mu}$, panel (c). } 
  \label{fig:lRPM}
\end{figure}

Next, a loop $(\vec{\mu}, \vec{\nu})$ has the {\em absorption property}, if for each $\Up$-boundary state $\vec{\nu}_i$ and $\Dn$-boundary $\vec{\mu}_j$ the following is true: $\vec{\mu} \in \Dn^* \vec{\nu}_i$ and 
$\vec{\nu} \in \Up^* \vec{\mu}_j$. Note that if the absorption property holds, the pairs 
$(\vec{\mu}, \vec{\nu}_i)$ and $(\vec{\mu}_j, \vec{\nu})$ form loops themselves. We call these the major sub-loops of $(\vec{\mu}, \vec{\nu})$. The absorption property is illustrated in Fig.~\ref{fig:lRPM}(a).

The {\em loop} return point memory ($\ell$RPM) property for the AQS transition graph can now be stated as 
follows:
\begin{itemize}
 \item [] {\em \bf $\ell$RPM}: The AQS transition graph has the $\ell$RPM) property, if {\em every} loop $(\vec{\mu}, \vec{\nu})$ has the absorption property. 
\end{itemize}
Note that the $\ell$RPM, as given above is defined as a property of the whole AQS transition graph. This is the case for the Preisach model, since {for the Preisach model} RPM is a direct consequence of NP, which in turn is a global dynamic property \footnote{In the case of sheared amorphous solids and other glasses, it is more useful to work with a local version of $\ell$RPM that is restricted to a given loop $(\vec{\mu}, \vec{\nu})$, as was done in \cite{mungan2019networks}. The  $\ell$RPM property for a loop can be stated as follows: 
A loop $(\vec{\mu}, \vec{\nu})$ has the $\ell$RPM property, if the following two conditions hold: (i) $(\vec{\mu}, \vec{\nu})$ has the absorption property and (ii) all major sub-loops of $(\vec{\mu}, \vec{\nu})$ possess the $\ell$RPM property. 
}.   

An immediate consequence of the $\ell$RPM property is that the $\Dn$-orbits off the $\Up$-boundary of a loop cannot cross, but must merge 
at or prior to reaching the lower endpoint $\vec{\mu}$. This is illustrated in panels (b) and (c) of Fig.~\ref{fig:lRPM}. An analogous 
result holds for the $\Up$-orbits off the $\Dn$-boundary of the same loop. In ref.~\cite{munganterzi2018} this has been called the 
nesting property, Proposition 3.5. With the help of the nesting property a series of results follow that we summarize next. 

\subsection{$(\mu, \nu)$ - reachable states} 
\label{subseq:munureachable}

With {each} loop $(\vec{\mu}, \vec{\nu})$ we can associate a set of stable states $\mathcal{R}_{(\vec{\mu}, \vec{\nu})} \subset \mathcal{S}$. 
{We refer to these states as $(\vec{\mu}, \vec{\nu})$-{\em reachable states}: these are} the states that can be reached from $\vec{\mu}$ by applying some sequence of $\Up$ and $\Dn$ operation such that the resulting intermediate states never leave the 
endpoints of the loop, {\em i.e.} transitions $\Up \vec{\nu}$ or $\Dn \vec{\mu}$ are not permitted. 
Since $\mathcal{S}$ is assumed finite, so must be $\mathcal{R}_{(\vec{\mu}, \vec{\nu})}$. 
Moreover, the $\ell$RPM property assures that for any $\vec{\sigma} \in \mathcal{R}_{(\vec{\mu}, \vec{\nu})}$, 
\begin{equation}
    \vec{\nu} \in \Up^* \vec{\sigma}, \quad \vec{\mu} \in \Dn^*\vec{\sigma},
    \label{eqn:localabsorption}
\end{equation}
which in turn implies {that $(\vec{\mu}, \vec{\nu})$-reachable states exit their loop via their endpoints}.  

{The set of $(\vec{\alpha}, \vec{\omega})$-reachable states are associated with the main hysteresis loop. We shall refer to them just 
as reachable states and denote this set by $\mathcal{R}$}. Since the endpoints of the loop $(\vec{\alpha}, \vec{\omega})$ are absorbing, property \eqref{eqn:localabsorption} is trivial in this case.  

{Referring to the example of Fig.~\ref{fig:Gen_defs}(b), the states
$\vec{\alpha}, \vec{\alpha_1}, \vec{\alpha}_2, \vec{\alpha}_3, \vec{\alpha}_4, \vec{\omega}, \vec{\omega_1}, \vec{\omega}_2, \vec{\omega}_3, \vec{\omega}_4$,  constitute the set of reachable states $\mathcal{R}$. Likewise, $\mathcal{R}_{(\vec{\mu},\vec{\nu})} = \{ \vec{\mu}, \vec{\nu} \}$ are the set of reachable states associated with the 2-loop $(\vec{\mu},\vec{\nu})$. 
}


\subsection{Standard partitioning of a loop into sub-loops}
\label{subsec:standard}

\begin{figure}[t!]
  \begin{center}
    \includegraphics[width = 3in]{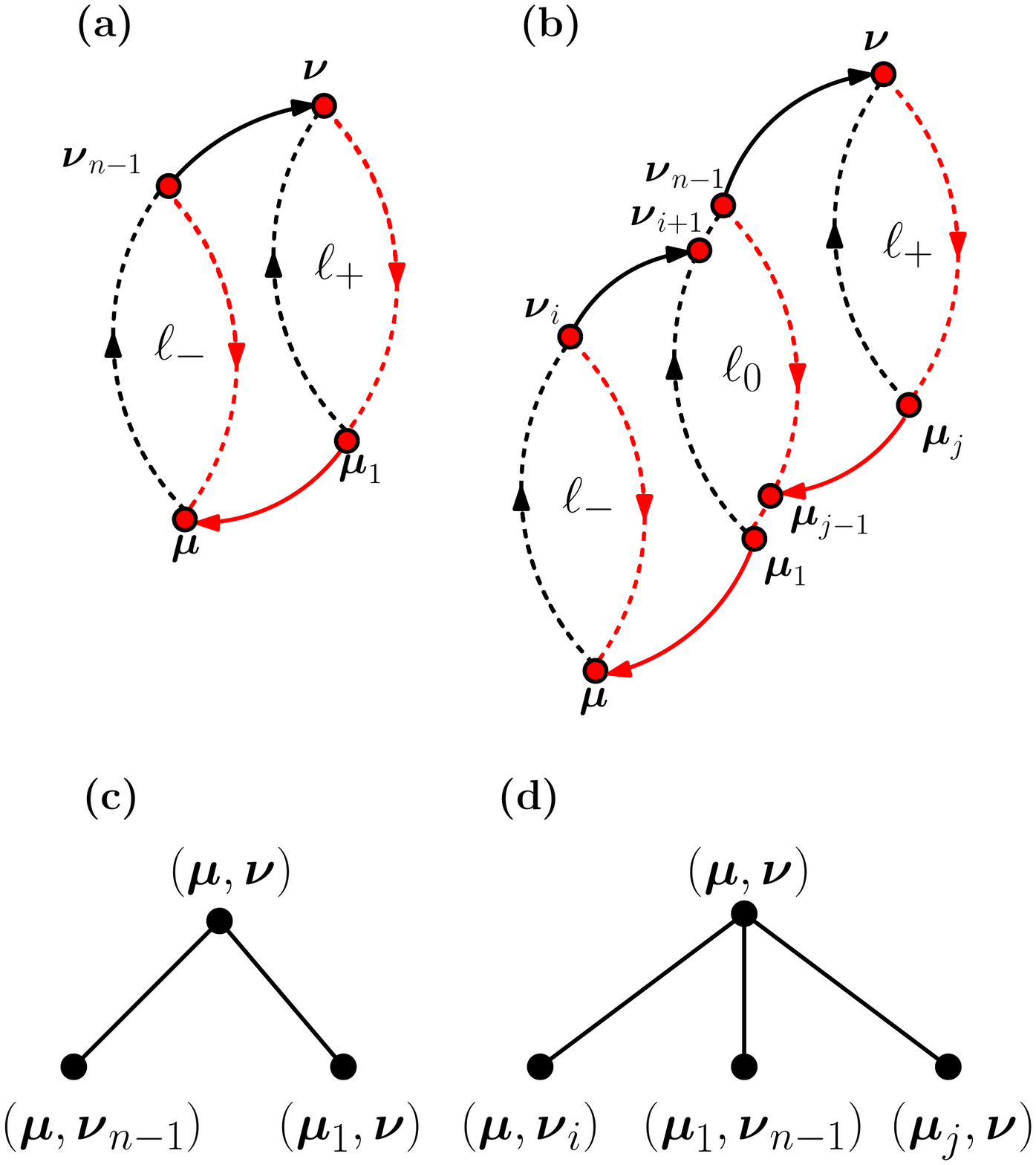} 
  \end{center}
  \caption{(color online) The two possibilities for the standard partitioning of a loop $(\vec{\mu}, \vec{\nu})$ into two (a) or three (b) 
  sub loops. The sub loops are marked as $\ell_-, \ell_+$ and $\ell_0$. Panels (c) and (d) depict the tree representation of 
  the partitioning of the parent loop $(\vec{\mu}, \vec{\nu})$ into two or three off-spring sub loops.
  } 
  \label{fig:CanonicalPart}
\end{figure}

{Given a loop $(\vec{\mu}, \vec{\nu})$ and its associated set of reachable states $\mathcal{R}_{(\vec{\mu}, \vec{\nu})}$, we say that a pair $(\vec{\kappa}, \vec{\lambda})$ of $(\vec{\mu}, \vec{\nu})$-reachable states  forms a {\em sub-loop} of $(\vec{\mu}, \vec{\nu})$, if 
$(\vec{\kappa}, \vec{\lambda})$ forms a loop.}
The $\ell$RPM property permits the decomposition of a loop $(\vec{\mu}, \vec{\nu})$ into two or more sub-loops. A particular way of doing this has been called {\em standard partitioning} and is illustrated in Fig.~\ref{fig:CanonicalPart}. Depending on whether the pair of states $\vec{\mu}_1$ and $\vec{\nu}_{n-1}$ forms a loop or not, $(\vec{\mu}, \vec{\nu})$ can be decomposed into two  or three loops, panels (a) and (b), respectively.
One can think of the decomposition as removing the transitions indicated by the solid arrows shown in Fig.~\ref{fig:CanonicalPart}(a) and (b). The $\ell$RPM property ensures that the state transition graph is thereby broken into disjoint components, meaning that no transitions from one component loop to another remain.     
The resulting component loops by definition possess $\ell$RPM and can therefore be partitioned in the same way. This partitioning procedure can be continued until all remaining components are singleton loop, {\em i.e.} loops whose lower and upper endpoint coincide.

{In the example of Fig.~\ref{fig:Gen_defs}(b), the standard partitioning of the main hysteresis loop $(\vec{\alpha},\vec{\omega})$ results in the three  sub-loops $(\vec{\alpha},\vec{\omega}_3)$, $(\vec{\alpha}_1,\vec{\omega}_4)$, and $(\vec{\alpha}_4,\vec{\omega})$.}

{Note that the  decomposition of the parent loop into component loops also furnishes a partition of the set of reachable states associated with the parent loop into the disjoint sets of reachable states associated with each of the component loops}.
Thus the standard partitioning procedure  amounts to a successively finer partition of the reachable states $\mathcal{R}_{(\vec{\mu}, \vec{\nu})}$ associated with the loop 
$(\vec{\mu}, \vec{\nu})$.

The standard partitioning procedure has a representation as a tree whose vertices are the loops and in which the offspring nodes are the components of the parent loop obtained under the standard partition procedure, as shown in Fig.~\ref{fig:CanonicalPart}(c) and (d). {An immediate consequence of this tree representation is that the transition graph formed by the set of 
reachable states $\mathcal{R}_{(\vec{\mu}, \vec{\nu})}$ of any loop $(\vec{\mu}, \vec{\nu})$ possessing the $\ell$RPM property must be planar \cite{munganterzi2018}. A similar result for the planarity of the AQS transition graph associated with  a certain class of dynamical systems has been obtained in  \cite{rachinskii2016realization}. } 

\subsection{Maximal loops}
\label{subsec:maxloops}

As is already apparent from Fig.~\ref{fig:CanonicalPart}(a) and (b), given two or more loops obeying the $\ell$RPM property, their endpoints can be connected so that a larger loop possessing $\ell$RPM is formed. If the state transition graph of an AQS automaton is such that every loop possesses the $\ell$RPM property, as is the case for the Preisach model, then, using the observation just made, it follows that every loop is a sub-loop of a unique largest loop. We call such loops {\em maximal loops}. {Formally, a loop $(\vec{\mu}, \vec{\nu})$ is called {\em maximal}, if (i) for any state $\vec{\nu'} \ne \vec{\nu}$ on the $\Up$-orbit of $\vec{\nu}$,  $\Dn^*\vec{\nu'}$ does not contain any  $(\vec{\mu},\vec{\nu})$-reachable state, and likewise, for any $\vec{\mu'} \ne \vec{\mu}$ of $\Dn^*\vec{\mu}$, $\Up^*\vec{\mu'}$ does not contain any $(\vec{\mu},\vec{\nu})$-reachable state. For example in Fig.~\ref{fig:Gen_defs}(c), the loops $(\vec{\mu}, \vec{\nu})$ 
and $(\vec{\alpha}, \vec{\omega})$ are maximal loops. Note that due to the absorption properties of its endpoints, \eqref{eqn:aoabsorption}, the main hysteresis loop always is a maximal loop.}  

In Ref.~\cite{munganterzi2018} we have introduced an algorithm that determines the maximal loop containing a given loop $(\vec{{\mu}}_0, \vec{{\nu}}_0)$:
\begin{itemize}
    \item[(A0)] Initialize {$(\vec{{\mu}}, \vec{{\nu}})$ as $(\vec{{\mu}}_0, \vec{{\nu}}_0)$} 
    \item[(A1)] Determine the largest $m$ that satisfies the condition
    \begin{equation}
    \vec{{\mu}} \in \Dn^{*} \left ( \Up^{m} \vec{{\nu}} \right ) \ ,    \nonumber
    \end{equation}
    and set $\Up^{m} \vec{{\nu}}$ {to} $\vec{{\nu}}$,
    \item[(A2)] Determine the largest $n$ that satisfies the condition
    \begin{equation}
    \vec{{\nu}} \in \Up^{*} \left ( \Dn^{n} \vec{{\mu}} \right ) \ ,    \nonumber
    \end{equation}
    and set $\Dn^{n} \vec{{\mu}}$ {to} $\vec{{\mu}}$,
    \item[(A3)] Repeat steps (A1) and (A2) until both $m$ and $n$ are zero and terminate.
\end{itemize}

The loop $(\vec{{\mu}}, \vec{{\nu}})$, obtained when the algorithm terminates, is the maximal loop containing $(\vec{{\mu}}_0, \vec{{\nu}}_0)$. It can be shown that the assignment of loops to maximal loops is unique, in the sense that the same maximal loop is reached via the above algorithm from any of its sub-loops.

As we have seen in Section \ref{subseq:munureachable}, the $\ell$RPM property permits us to associate with any loop $(\vec{{\mu}}, \vec{{\nu}})$ {a set of $(\vec{\mu}, \vec{\nu})$-{\em reachable states},  $\mathcal{R}_{(\vec{\mu}, \vec{\nu})}$.} Thus the set of maximal loops via their associated set of reachable states furnishes a partition of $\mathcal{S}$: each state $\vec{\sigma} \in \mathcal{S}$ belongs to exactly one maximal loop $(\vec{{\mu}}, \vec{{\nu}})$. 
We shall call the maximal loops consisting of a single state $\vec{\sigma}$, singleton maximal loops. For such loops 
$\vec{\mu}_0 = \vec{\nu}_0 = \vec{\sigma}$, and the algorithm terminates right away. 

The utility of the maximal loops lies in the local absorption property, \eqref{eqn:localabsorption}. Thus a sequence of transitions connecting two  states belonging to different maximal loops must be such that the maximal loops involved are left through their  endpoints. This in turn leads to a coarse-grained description of the dynamics by means of a condensed state transition graph where each vertex is a maximal loop, the inter-loop state transition graph \cite{munganterzi2018}. The topology of the inter-loop transition graph provides useful information about the dynamics, such as the length of transients upon cyclic forcing. We now turn to the structure of maximal loops in the Preisach model. 

\section{Maximal loops of the Preisach Model}
\label{sec:preisachmax}

In this section we discuss the maximal loops of the Preisach model.
We first define the size of a loop as the number of state changes that occur as the loop is traversed from one endpoint to the other. Since by property (P) each $\Up$ and $\Dn$ step changes the state of only one hysteron, the size of a loop is also equal to the number of hysterons that change their state as the loop is traversed. Therefore, the endpoints of a given loop $(\vec{{\mu}}, \vec{{\nu}})$ with size $j$ are mapped into each other according to
\begin{equation}
    \vec{\mu} = D^j \vec{\nu} \ \ \ \mbox{and} \ \ \ \vec{\nu} = U^j \vec{\mu}.
\end{equation}
The maximal loop that contains a given loop $(\vec{{\mu}}, \vec{{\nu}})$ is found by the algorithm introduced in Section \ref{subsec:maxloops}. 

Recall the observation made in Section \ref{subsec:rhostable} that there exists a subset of stable states $\mathcal{S}_\rho$, whose stability is a direct consequence of the permutation $\vec{\rho}$, defined by \eqref{eqn:rhoorder}, which orders the switching fields $F^-_i$. We have called such states $\rho$-stable. In particular,  we showed that all two-loops are $\rho$-stable, property (P1) and Eq.~\eqref{eqn:S2subsetSrho}.   By property (P), arbitrary sequences of $\Up$ and $\Dn$ operations applied to a state $\vec{\sigma}$ in $\mathcal{S}^{(2)}$ must lead to some stable state $\vec{\sigma'} \in \mathcal{S}_\rho$. Moreover, as a result of the maximal loop property, any pair of states forming a two loop is part of some maximal loop. 
Denote by {$\mathcal{M} \subset \mathcal{S}$ the set of states that are reachable states of some non-singleton maximal loop}. Note that {$\mathcal{M}$} can in principle contain states that are not $\rho$-stable. For the Preisach model this turns out  not to be the case however,  and we have:
\begin{itemize}
    \item[(P2)] Any state {that is a reachable state of some} non-singleton maximal loop is also the endpoint of some two-loop so that 
    \begin{equation}
        \mathcal{S}^{(2)} = \mathcal{M}. 
    \end{equation}
\end{itemize}
In order to prove (P2), we first assume that a state $\vec{\sigma}$ is \emph{not} an endpoint of a two-loop. Then, we show that the state $\vec{\sigma}$ cannot be a part of a non-singleton maximal loop. Recall that a singleton loop is a loop where the two endpoints coincide, so we can apply the maximal loop finding algorithm to $\vec{\sigma}$ and find the endpoints of the maximal loop it is part of. In step A0 of the algorithm, we therefore have $ \vec{\sigma}= \vec{\mu}_0 = \vec{\nu}_0 $. Since we are assuming that $\vec{\sigma}$ is not an endpoint of a two-loop, we have  $\Dn \Up\vec{\sigma} \neq \vec{\sigma}$ and $\Up\Dn \vec{\sigma} \neq \vec{\sigma}$. Using {(P)}, we can write these two conditions as
\begin{equation}
    \vec{\sigma} \not\in \Dn^* (\Up \vec{\sigma}) \ \ \ \mbox{and} \ \ \ \vec{\sigma} \not\in \Up^* (\Dn \vec{\sigma}).
\end{equation}
These conditions together with the absorption property underlying $\ell$RPM ({\em} cf. Section \ref{sec:rpm}), 
imply that the largest integers $m,n$ satisfying the conditions in steps A1 and A2 of the algorithm are zero. Hence the algorithm terminates right away and the state $\vec{\sigma}$ is  a singleton maximal loop. A state that is not an endpoint of a two-loop cannot be a part of a non-singleton maximal loop and therefore property (P2) holds. 

From (P2) and the $\rho$-stability of two-loops, property (P1), it immediately follows that
\begin{itemize}
    \item[(P3)] The states of all non-singleton maximal loop are $\rho$-stable{:}
    \begin{equation}
      \mathcal{M} \subset \mathcal{S}_{\rho}.  
    \end{equation}
\end{itemize}
In other words, all states that are part of some loop which involves two or more states, are necessarily $\rho$-stable. Consequently, the topology of a {\em structurally} large part of the state transition graph, namely the transition among states that are part of non-trivial loops, is determined entirely by the permutation $\vec{\rho}$.  

To summarize,  
\begin{equation}
  \mathcal{M} \subset \mathcal{S}_{\rho} \subset \mathcal{S},     
\label{eqn:S2subsetSrhorelation}
\end{equation}
and the only stable states that are not $\rho$-stable are singleton maximal loops.
{By (P3), $\mathcal{S}_{\rho} \setminus \mathcal{M}$, is the set of  singleton maximal loops that are $\rho$-stable. } 
{It turns out that this set is empty and thus 
\begin{equation}
    \mathcal{M} = \mathcal{S}_{\rho}.
\end{equation}
The Preisach model does not have $\rho$-stable singleton maximal loops \footnote{The proof of this statement is not difficult, but beyond the scope of this article. We sketch out the key ideas leading to it. We consider realizations of $\rho$-compatible switching fields $F^\pm_i$, {\em i.e.} realizations that satisfy \eqref{eqn:tipF}, \eqref{eqn:Fplus_order}, and
\eqref{eqn:rhoorder}. Among these, the realizations for which the $F^-_i$ take their largest possible value relative to some fixed values of the $F^+_i$, will have the following properties: (i) all states of this realization are $\rho$-stable, $\mathcal{S}= \mathcal{S}_{\rho}$, and  (ii),  $\mathcal{S}^{(2)} = \mathcal{S}_{\rho}$, from which it follows that 
$\mathcal{S}_{\rho}$ does not  contain singleton maximal loops, so that $\mathcal{M} = \mathcal{S}_\rho$.}.}

To give an example of a singleton maximal loop and its stability, consider  the state $\vec{\sigma}_1 = (-1,+1,-1,-1,-1)$ in Fig.~\ref{fig:Gen_defs}(b). With $\vec{\rho} = (35214)$ and using \eqref{eqn:rhoorder}, it is readily checked that $F^+[\vec{\sigma}_1] = F^+_1$, while $F^-[\vec{\sigma}_1] = F^-_2$. The permutation $\vec{\rho}$ is compatible with an ordering of the switching fields as $F^+_1 < F^-_2 < F^+_2$, which in turn would imply that $\vec{\sigma}_1$ is not a stable state. Thus specifying the ordering permutation $\vec{\rho}$ does not suffice to guarantee the stability of this state. We will not pursue the properties of singleton maximal loop states further in this paper. 

\section{The structure of the Preisach graph}
\label{sec:preigraph}

We are primarily interested in the structure of the state-transition graph on $\mathcal{R}$, {the set of reachable states associated with  
the main hysteresis loop $(\vec{\alpha}, \vec{\omega})$}.
{Since $(\vec{\alpha}, \vec{\omega})$ is a maximal loop, it follows from property {(P3)} of Section \ref{sec:preisachmax}  that all states of $\mathcal{R}$ are $\rho$-stable}. 
Thus the permutation $\vec{\rho}$ determines the entire transition graph on $\mathcal{R}$. We shall call this graph the {\em Preisach graph}. Using the $\ell$RPM property defined in Section \ref{sec:rpm}, our main goal in this section is to work out the topology of this  graph and show how it is determined by $\vec{\rho}$. We will then address the reverse problem of how, given an unlabeled Preisach graph, the permutation  $\vec{\rho}$ can be inferred from its topology. 

We denote the $\Up$- and $\Dn$-boundary states of $(\vec{\alpha}, \vec{\omega})$  respectively as $\vec{\omega}_i = \Up^i \vec{\alpha} $ and $\vec{\alpha}_i = \Dn^{L-i} \vec{\omega}$, with $i = 0, 1, 2, \ldots, L$. {This labeling  is illustrated in the example of Fig.~\ref{fig:Gen_defs}(b).}  

Having ordered the switching fields $F^+_i$ as in \eqref{eqn:Fplus_order}, and using property (P), the $\Up$-boundary states of the loop $(\vec{\alpha}, \vec{\omega})$ are given in terms of the sets $I^+[\vec{\omega}_i]$, {\em cf.} \eqref{eqn:iplusdef}, as
\begin{equation}
    I^+[\vec{\omega}_i]  = \left \{ \begin{array}{cc}
      \emptyset, & i = 0, \\
      \{1, 2, \ldots, i\}, & 0 < i \le L.
 \end{array} \right.    
\end{equation}
Likewise, for the configurations associated with the $\Dn$-boundary states $\vec{\alpha}_i = \Dn^{L - i} \vec{\omega}$, we find from 
\eqref{eqn:iminusdef} that 
\begin{equation}
    I^-[\vec{\alpha}_i]  = \left \{ \begin{array}{cc}
      \emptyset, & i = L, \\
      \{\rho_1, \rho_2, \ldots, \rho_{L -i} \}, & 0 \le i < L.
 \end{array} \right.    
\end{equation}

As discussed in Section \ref{subsec:standard} and illustrated in Fig.~\ref{fig:CanonicalPart}(a)-(b)), 
the $\ell$RPM property permits a partitioning of the loop $(\vec{\alpha}, \vec{\omega})$ 
into two or three sub-loops. We have called this procedure the standard partitioning of the loop.   
{For the Preisach model} whether the standard partitioning results in two or three loops turns out to depend on the position of the element $L$ in the ordering 
permutation $\vec{\rho}$. 

\begin{figure*}[t!]
  \begin{center}
    \includegraphics[width =\textwidth]{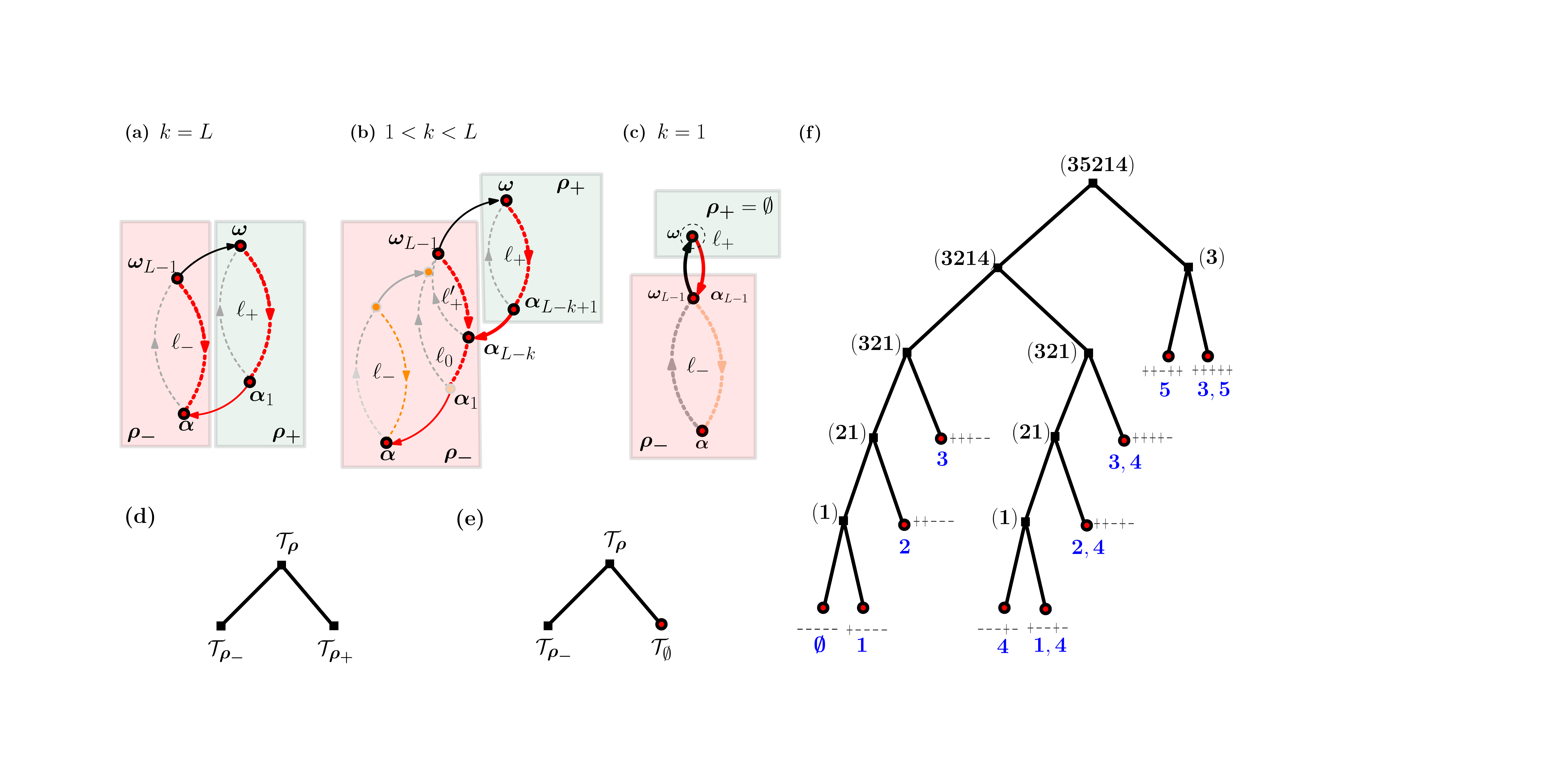} 
  \end{center}
  \caption{(color online) (a) - (c): Preisach partitioning of the main hysteresis loop $(\vec{\alpha}, \vec{\omega})$. This partition is an adaptation of 
  the standard partition, {\em cf. } Fig.~\ref{fig:CanonicalPart} (a) and (b), and always results in a pair of loops: the ''left`` and 
  ''right`` loops, labeled as $\vec{\rho_-}$ and $\vec{\rho_+}$ and also highlighted, respectively  by boxes with red and green backgrounds. The ''left`` loop has endpoints 
  $(\vec{\alpha}, \vec{\omega}_{L-1})$ and is a consolidation of the loops $\ell_-$ and $\ell_0$ of the standard partitioning. 
  The ''right`` loop has endpoints $(\vec{\alpha}_{L-k+1}, \vec{\omega})$, where $k$ is the element of the ordering permutation 
  $\vec{\rho}$ for which $\rho_k = L$. The special cases $k = L$ and $k = 1$ are depicted in panels (a) and (c). The middle loop of the 
  standard partition is only present when $k \ne L$. 
  The ''left`` and ''right`` sub-loops are the hysteresis loops of Preisach systems with $L-1$, resp. $k$, hysterons and are generated by the permutations   $\vec{\rho_-}$ and $\vec{\rho_+}$, as defined in \eqref{eqn:rhominusdef} and \eqref{eqn:rhoplusdef}.(d) The 
  Preisach partitioning of a parent loop $\vec{\rho}$ into two off-spring loops $\vec{\rho_-}$ and $\vec{\rho_+}$ can be represented in terms of a tree, shown in (d) and (e) and corresponding to the cases (a)-(b) and (c), respectively. When $k = 1$, the "right" loop  has a single state, and $\vec{\rho}_+ = \emptyset$, {\em cf.} \eqref{eqn:rhoplusdef}. 
  (f) Tree representation of the Preisach partitioning of the loop $(\vec{\alpha}, \vec{\omega})$  of  Fig.~\ref{fig:Gen_defs}(b). The loop is generated by the permutation $\vec{\rho} = (35214)$. The parent-child relation as given by (d) and (e), with the {Preisach graph} constituting the root of the tree. Nodes of the tree shown that are not leaves are depicted by black boxes. They correspond to intermediate loops and the labels next to these indicate the permutations that generate them. The leaves of the tree are the states associated with the {Preisach graph}. They turn out to be in $1$-to-$1$ correspondence with the set of increasing subsequences contained in $\vec{\rho}$. The labels below each leave node establish this correspondence. 
  Refer to text for further details.
  } 
  \label{fig:PreisachStandardPart}
\end{figure*}

Let $k$ be the element of $\vec{\rho}$ that is mapped to $L$, so that 
$\rho_k = L$. As we will see shortly, a partition into two loops occurs only if $k = L$ and hence $\rho_L = L$. In this case 
the loops are labeled as $\ell_-$ and 
$\ell_+$, with endpoints $(\vec{\alpha}, \vec{\omega}_{L-1})$ and $(\vec{\alpha}_1, \vec{\omega})$, respectively. This is shown 
in Fig.~\ref{fig:PreisachStandardPart}(a). In all other cases the partition yields three loops which we label as 
$\ell_-, \ell_0$, and $\ell_+$, with 
the center loop $\ell_0$ having endpoints $(\vec{\alpha}_1, \vec{\omega}_{L-1})$. Fig.~\ref{fig:PreisachStandardPart}(b) depicts the 
three-way partition of the {Preisach graph} when  $1 < k < L$. 

For the Preisach model it turns out to be more convenient to define a binary partition where the ``left'' loop is the consolidation 
of $\ell_-$ and $\ell_0$ (if it exists), and the ''right`` loop is the loop $\ell_+$. {In Fig.~\ref{fig:PreisachStandardPart} (a) - (c), these partitions are highlighted by boxes shaded in light red and green and labeled as $\vec{\rho}_-$ and $\vec{\rho}_+$, respectively.} Regardless of whether the standard partition would have led to two or three loops, 
the ''left`` loop of the binary partition will always have endpoints $(\vec{\alpha}, \vec{\omega}_{L-1})$. The ''right`` loop has 
upper endpoint $\vec{\omega}$. We claim that its lower endpoint is given by the boundary state $\vec{\alpha}_{L - k + 1}$, with 
$k$ such that $\rho_k = L$. 

To see this, consider the configurations $\vec{\omega}_{L-1}$ and $\vec{\omega}_{L} = \vec{\omega}$ on the $\Up$-boundary of the loop $(\vec{\alpha}, \vec{\omega})$. Recall that the $\ell$RPM property implies the nesting property of orbits off the boundary of a 
loop, as illustrated in Fig.~\ref{fig:lRPM} (b) and (c). Consequently, the two orbits $\Dn^* \vec{\omega}_{L-1}$ 
and  $\Dn^* \vec{\omega}$ must either merge at $\vec{\alpha}$, or they must merge further "upstream" at some state that lies on 
the $\Dn$-boundary of the loop $(\vec{\alpha}, \vec{\omega})$. We claim that the merging occurs at the $\Dn$ boundary 
state $\vec{\alpha}_{L-k}$, with $k$ being again the element of $\vec{\rho}$ for which $\rho_k = L$. In fact, note that for each $i$ with  $0 \le i < k$, the configurations $\Dn^{i} \vec{\omega}_{L-1}$ and $\Dn^{i} \vec{\omega}$ differ only by the state of the $L$th hysteron. Moreover, 
since $\rho_k = L$, it follows that $\Dn^{k} \vec{\omega} = \Dn^{k - 1} \vec{\omega}_{L-1}$. This means that 
the orbits $\Dn^* \vec{\omega}_{L-1}$ and $\Dn^* \vec{\omega}$ merge at the state $\vec{\alpha}_{L-k}$,  
as illustrated in Fig.~\ref{fig:PreisachStandardPart}(a) - (c). We have thus shown that a {Preisach graph} can always be partitioned into 
a 'left'' and ''right`` loop, with its endpoints given by $(\vec{\alpha}, \vec{\omega}_{L-1})$ and   
$(\vec{\alpha}_{L -k + 1}, \vec{\omega})$, respectively. 
{The following observations are an immediate consequences of this result.} 

First, by the $\ell$RPM property applied to the ''left`` loop,  
the pair of states $(\vec{\alpha}_{L-k}, \vec{\omega}_{L-1})$ must also form a loop. Let us denote this loop as 
$\ell'_0$. For $k = L$ this loop coincides with the loop 
$\ell_-$ of Fig.~\ref{fig:PreisachStandardPart}(a), while for $k = 1$, $\ell'_0$ is the singleton loop consisting of 
the state $\vec{\omega}_{L - 1} = \vec{\alpha}_1$, as shown in Fig.~\ref{fig:PreisachStandardPart}(a). The generic case is depicted in Fig.~\ref{fig:PreisachStandardPart}(b). Note that regardless of $k$, the loops $\ell'_0$ and $\ell_+$ are always isomorphic, since their endpoints and hence all other states associated with the 
two loops differ only by the state of the $L$th hysteron. 

Second, and more importantly, the ''left`` and ''right`` loops can be regarded as the main hysteresis loops 
of two smaller Preisach systems with $L-1$ and $k - 1$ hysterons, respectively. 
Specifically, the 
transition graph associated with the ''left`` 
loop $(\vec{\alpha}, \vec{\omega}_{L-1})$ is the {Preisach graph of a system} with $L-1$ hysterons. These are the hysterons $1, 2, \ldots, L - 1$ of the parent 
Preisach system. The ordering permutation $\vec{\rho_-}$ associated with this subsystem is thus obtained from 
$\vec{\rho}$ by removal of the entry for $L$, so that
\begin{equation}
    \vec{\rho_-} = \left \{ \begin{array}{cc}
    (\rho_2, \ldots, \rho_L), & k = 1, \\
    (\rho_1, \ldots, \rho_{k-1}, \rho_{k+1}, \ldots, \rho_L), & 1 < k < L, \\
    (\rho_1, \ldots, \rho_{L-1}), & k = L, 
    \end{array}
    \right.
    \label{eqn:rhominusdef}
\end{equation}
where $k$ is the index for which $\rho_k = L$. Define also  
$\vec{\rho_+}$ as 
\begin{equation}
    \vec{\rho_+} = \left \{ \begin{array}{cc}
    \emptyset, & k = 1, \\
    (\rho_1, \rho_2, \ldots, \rho_{k-1}), & 1 < k \le L.
    \end{array}
    \right.
    \label{eqn:rhoplusdef}
\end{equation}
We see that for $k > 1$, $\vec{\rho_+}$ contains the first $k-1$ elements of $\vec{\rho}$. By the argument leading to the binary partition 
of a Preisach graph, it follows that the ''right`` loop is isomorphic to a 
{Preisach graph} generated by a system of $k - 1$ hysterons, labeled in terms 
of the corresponding hysterons of the parent system by
the elements of $\vec{\rho_+}$. By our ordering convention \eqref{eqn:Fplus_order} for the switching fields, the sequence in which each hysteron of the subsystem changes 
its state when moving from the lower to upper endpoint, is given by the ordering of their labels from smallest to largest. The permutation $\vec{\rho_+}$ therefore 
prescribes again the sequence of hysteron state changes, as we move back from the upper to the lower endpoint. 
When $k = 1$, the 'right'' loop is a singleton loop, {\em cf. } Fig.~\ref{fig:PreisachStandardPart}(c). Since a Preisach system with one hysteron 
already has two states, it is convenient to interpret such a singleton loop as corresponding to a Preisach system with no hysterons at all. 
We assign the empty set $\emptyset$ as its ordering permutation.

We have thus shown how the {Preisach graph} associated with $\vec{\rho}$ can be partitioned into two sub-loops that in turn are the 
Preisach graphs generated by $\vec{\rho}_-$ and $\vec{\rho}_+$. The Preisach partition in effect removes the $L$th hysteron, resulting in two sub-systems with $L - 1$ and $k - 1$ hysterons, with ordering permutations given by \eqref{eqn:rhominusdef} and \eqref{eqn:rhoplusdef}.
Fig.~\ref{fig:PreisachStandardPart} (d) and (e) depict the parent-child relation induced by this partition 
for the cases $k > 1$ and $k = 1$, respectively.   
The ``left'' and ``right'' {Preisach graphs} in turn can be partitioned in 
a similar manner and this procedure can be continued, until all loops are singleton loops and hence cannot be further partitioned. 

Fig.~\ref{fig:PreisachStandardPart} (f) depicts the Preisach partition of the loop $(\vec{\alpha},\vec{\omega})$ in 
Fig.~\ref{fig:Gen_defs} (b).  By virtue of the distinction between ``left'' and ``right'' sub-loops, this is again an ordered tree. 
The non-leaf nodes of this tree correspond to Preisach subsystems with one or more hysterons and are labeled by the corresponding 
permutations generating these. The root node is the main hysteresis loop with ordering permutation $\vec{\rho} = (35214)$ so that $k = 2$. Its Preisach partition thus 
yields the left and right off-spring loops that are generated by the permutation $\vec{\rho}_- = (3214)$ and $\vec{\rho}_+ = (3)$. The leaves of the tree are the $10$ states constituting the main hysteresis loop and all of its sub-loops. We have identified these states by their hysteron configurations.

We conclude this section by showing how, given an unlabeled {Preisach graph}, one can infer from its topology the ordering permutation 
$\vec{\rho}$ generating it. The endpoints of the loop are easily identified. Counting the number of transitions from the lower 
to upper endpoint, we obtain the number of hysterons $L$. Denote the $\Up$-boundary states of the loop as $\vec{\omega}_i = \Up^i \vec{\alpha}$, 
with $i = 0, 1, 2, \ldots, L$. Consider next the $\Dn$-orbits off a pair of successive $\Up$-boundary states $\vec{\omega}_{i-1}$ and 
$\vec{\omega}_i$. By the $\ell$RPM property these orbits must merge by the time the lower endpoint is reached. For $i = 1, 2, \ldots, L$, 
let $k_i$ be the number 
of transitions after which the $\Dn$-orbit off $\vec{\omega}_i$ merges with the $\Dn$-orbit from $\vec{\omega}_{i-1}$. The permutation 
$\vec{\rho}$ is then obtained from the set of $L$ integers $k_i$ by the following procedure. Initialize $\vec{\rho}$ as $L$ empty slots. Proceeding in decreasing   
order, $i = L$, $i = L - 1$ {\em etc}, placing the symbol $i$ in the $k_i$th empty slot counted from the left. Upon completion, 
the slots contain the permutation $\vec{\rho}$. 

A related question is whether, given the transition graph associated with an $\ell$RPM loop, one can decide whether it is a Preisach 
graph or not. A relative straight forward procedure is to infer, if possible, the permutation $\vec{\rho}$ from the main hysteresis loop {$(\vec{\alpha}, \vec{\omega})$}, as described above. If this is possible, 
perform next the Preisach partition and infer from the topology of the ``left'' and ``right'' loop the corresponding  
permutations. If these are related to $\vec{\rho}$ as prescribed by \eqref{eqn:rhominusdef} and \eqref{eqn:rhoplusdef}, then this is 
a valid partition. If this validation procedure can be recursively repeated on the sub-loops without any inconsistencies, the given loop 
is a {Preisach graph}. 

Observe that by the isomorphism of the sub-loops $\ell'_0$ and $\ell_0$ arising in the course of the partition, {\em see}  
Fig.~\ref{fig:PreisachStandardPart}(b), the Preisach graph contains many repeated motifs of various sizes, since such isomorphisms 
persist at all levels of the partitioning. Thus the transition graph associated with a Preisach graph has a highly distinct topology.

\section{Statistics}
\label{sec:statistics}

In the following two-subsection we provide combinatorial results for the size-distribution of maximal loops and the number of reachable states associated with the main hysteresis loop. Denote by $\Pi_L$ the set of permutations of $L$ elements. We shall assume that the ordering permutation $\vec{\rho}$ is drawn at random and uniformly from $\Pi_L$. {For many applications, such an assumption is not realistic, and the distribution from which the switching fields are drawn will in general not result in a uniform distribution of $\vec{\rho}$. However, it is possible that the asymptotic forms of the results, which are obtained from saddle point approximations in the limit that $L$ becomes large, are robust to changes in the  distribution of $\vec{\rho}$. 

The calculation of the mean number of reachable states follows a divide-and-conquer approach that utilizes the $\ell$RPM property of the Preisach model. Such an approach can be applied to other models exhibiting $\ell$RPM, such as the toy model of depinning \cite{KMShort13,KMLong13}, and perhaps the random-field Ising model \cite{Sethna93}. This section therefore also serves as an example for how to use the $\ell$RPM property in a combinatorial setting. }   

\subsection{The size-distribution of maximal loops}

Let $(\vec{\mu},\vec{\nu})$ be {the} endpoints a maximal loop. Then, the maximal loop finding algorithm (A1), (A2) and (A3) imposes three conditions on {$\vec{\mu}$ and  $\vec{\nu}$:}
\begin{itemize}
\item[(B1)] $U^j \vec{\mu}$ = $\vec{\nu}$ and $D^j \vec{\nu}$ = $\vec{\mu}$,
\item[(B2)] $D^{j+1}U\vec{\nu}\neq\vec{\mu}$,
\item[(B3)] $U^{j+1}D\vec{\mu}\neq\vec{\nu}$.
\end{itemize}
Recall that the size $j$  of a maximal loop $(\vec{\mu}, \vec{\nu})$ is defined to be the number of hysterons that change their states as the loop is traversed. The first condition sets $\vec{\mu}$ and $\vec{\nu}$ as endpoints of a loop and the last two conditions ensure that the loop is a maximal loop. 

We will determine first the number $N_{L, j}$ of all possible maximal loops with size $j$ for given system size $L$. We will do this by summing over all possible $\vec{\rho}$ permutations. First, determine $N_{L, j}$ for the maximal loops that have $i$ hysterons in the configuration $+1$. We shall call the quantity $i$ the \emph{level}. The total number of maximal loop with size $j$ and level $i$ will be denoted by $N_{L, j, i}$. Then, by summing over $i$, $N_{L, j}$ will be determined.
 
Let us assume $(\vec{\mu}, \vec{\nu})$ is a maximal loop with size $j$ and level $i$. $(\vec{\mu}, \vec{\nu})$ must satisfy the conditions~(B1), (B2) and (B3). The condition of being a loop (B1) is now restated as a condition on the permutation $\rho$. During the evolution from $\vec{\mu}$ to $\vec{\nu}$ the first $j$ hysterons in state $-$ flip to $+$. If the same $j$ hysterons revert their states when the $\Dn^{j}$ operator is applied to $\vec{\nu}$, then $\vec{\mu}$ and $\vec{\nu}$ are the endpoints of a loop. The condition of having $\Dn^j\vec{\nu} = \vec{\nu}$ is that the $j$ positive hysterons that flipped back lie to the left of the initial $i$ hysterons in permutation $\rho$. In other words, we seek a permutation in which the $j$ hysteron labels come before the $i$ hysterons with no further restriction on the remaining $L-(i+j)$ hysterons. Therefore, the number of permutations that satisfy (B1) for  loops with level $i$ is
\begin{equation}
   N_{L, j, i}^{({\rm B1})} = \binomial{L}{i+j} i!j!(L-(i+j))!,
     \label{eqn:Condition-i}
\end{equation}
where the binomial coefficient gives the number of possible choice of $(i+j)$ hysterons, $i!$ gives the permutations of hysterons in state $+$ configurations, $j!$ gives the permutations of hysterons that revert their states as the loop is traversed, and $(L-(i+j))!$ gives the possible permutations of the remaining hysterons.

The number of permutations that satisfy the condition (B2) is determined similarly. First we find the number of loops with length $j+1$ and subtract this from the expression for the number of loops with size $j$. The result becomes,
\begin{equation}
     N_{L, j, i}^{({\rm B1})+({\rm B2})} = N_{L, j+1, i}^{({\rm B1})} - N_{L, j, i}^{({\rm B1})}
    = \frac{i\,L!\,i!\,j!}{(i+j+1)!},
    \label{eqn:Condition-ii}
\end{equation}
and it gives the number of permutations that satisfy both conditions (B1) and (B2). The ratio of expression~(\ref{eqn:Condition-ii}) to expression~(\ref{eqn:Condition-i}) $\frac{i}{i+j+1}$ gives us the ratio of loops, with size $j$ and level $i$ that do not extend with $\Up$-transition (via step (A1) of the maximal loop finding algorithm). Then, the ratio of loops of size $j$ and  level $L - (i+j)$ that do not extend with $\Dn$-transition is given by
\begin{equation}
   \frac{ N_{L, j, L-(i+j)}^{({\rm B1})+({\rm B3})}}{ N_{L, j, L-(i+j)}^{({\rm B1})}} =  \frac{L -(i+j)}{L - i + 1}. 
\end{equation}
Finally, multiplying this ratio with expression in \eqref{eqn:Condition-ii} we determine the number of possible permutations that satisfy conditions (B1), (B2) and (B3),
\begin{equation}
     N_{L, j, i} = \frac{i\,[L -(i+j)]}{L - i + 1} \frac{L!\,i!\,j! }{(i+j+1)!} \, .
\end{equation}

For a given system system size $L$, there are $L!$ different permutation $\rho$ which gives us $L!$ transition graphs. For a given system size $L$, among all $L!$ permutations, the number of maximal loops with size $j$ is therefore
\begin{eqnarray}
 N_{L, j} &= \displaystyle\sum_{i=1}^{L-j-1}  \binomial{L}{i} N_{L, j, i} \phantom{displaystyle\sum_{i=1}^{L-j-1} j j j  k k } \nonumber \\ 
 &=
 \displaystyle\sum_{i=1}^{L-j-1} i \, 
 \frac{L - i-j}{L-i+1} 
 \frac{L!\, i!\,j!}{(i+j + 1)!} 
 \binomial{L}{i},
 \label{eqn:dist}
\end{eqnarray}
where the binomial term gives the number of ways to choosing $i$ hysterons.

\begin{figure}[t!]
  \begin{center}
    \includegraphics[width =3in]{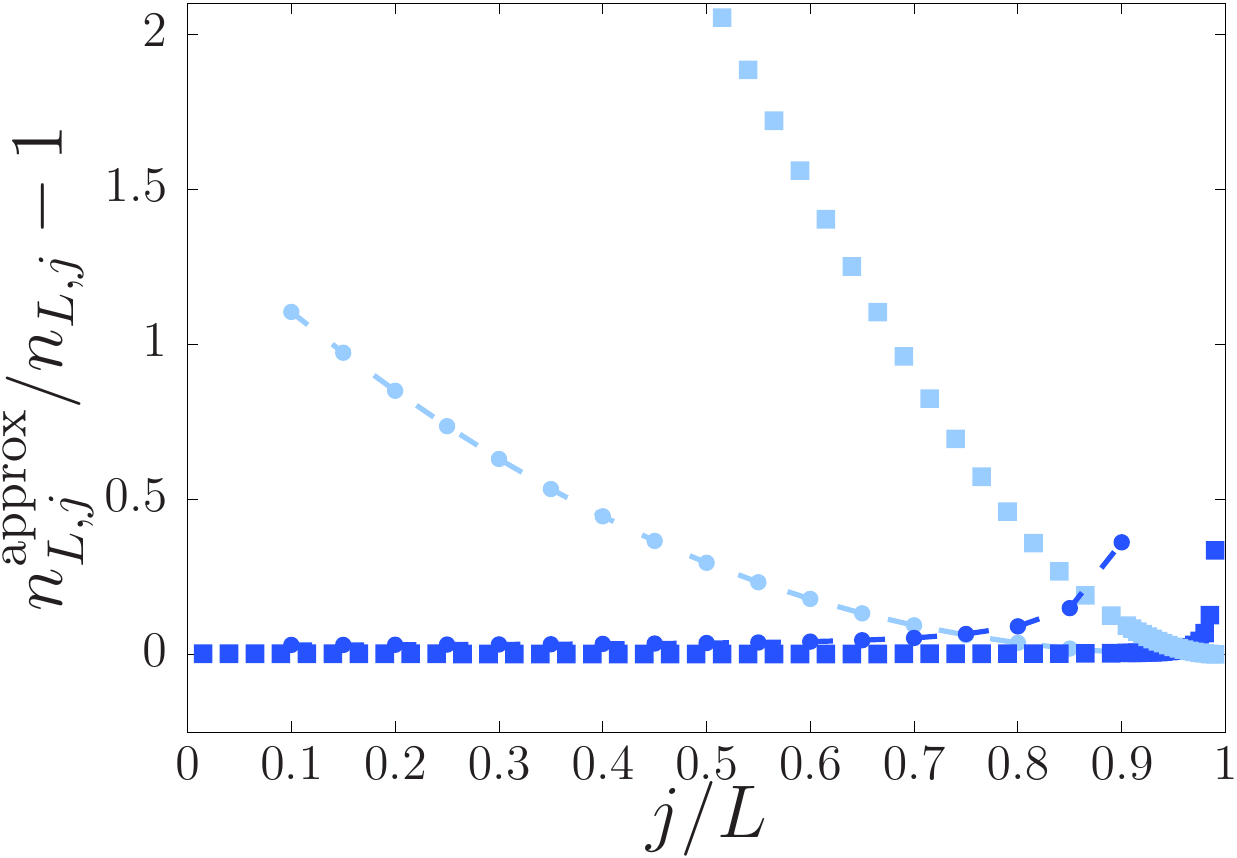} 
  \end{center}
  \caption{(color online) Ratios of the maximal loop distribution  Eqn.~(\ref{eqn:dist}) and its approximations, Eqn.~(\ref{eqn:approx-1}) (dark blue symbols) and Eqn.~(\ref{eqn:approx-2}) ({light blue symbols}), for  values of $L=20$ and $L=200$, distinguished by {circles} and {boxes}, respectively.
 } 
  \label{fig:dist}
\end{figure}

We are interested in the large $L$ limit of the the sum given in Eqn.~(\ref{eqn:dist}). In order to determine an asymptotic approximation, we first rescale the distribution $N_{L, j}$ as 
\begin{equation}
    n_{L,j} = \frac{(L+j+2)!}{L!\ L!\,j!} N_{L, j},
\end{equation}
which after some rearrangements becomes
\begin{equation}
    n_{L,j} = \displaystyle\sum_{i=1}^{L-j-1} 
    i\,(L-j-i) \binomial{L+j+2}{i+j+1}.
    \label{eqn:dist-2}
\end{equation}
Observe that the terms in the sum are symmetric around $i=(L-j)/2$ which also gives the largest binomial coefficient. Approximating the sum by an integral and making a saddle point approximation, we find
\begin{align}
    n_{L, j} &= 2^{L+j} (L-j)^2 
     \Big[\frac{L^2 - 6Lj -3 j^2}{2(L^2 - j^2)} \erf(t_0)  \nonumber\\
    & \phantom{=2^{L+j} (L-j)^2 
     \Big[ }
    + \frac{L^2 + 6Lj + j^2}{L^2 - j^2} \frac{t_0}{\sqrt{\pi}} e^{-t_0^2}
    \Big],
    \label{eqn:approx-1}
\end{align}
where $\erf(t)$ is the error function, while $t_0$ is the limit of the integral being  approximated and is given by 
\begin{equation}
    t_0 = \frac{1}{2}\sqrt{\frac{2}{L+j+2}} (L-j).
\end{equation}
For large $L$, Eqn.~(\ref{eqn:approx-1}) turns out to be a very good approximation of $n_{l,j}$ for the large systems as long as $j/L$ smaller than $\sim 0.9$. For values of $j/L$ close to $1$, the sum can no longer be approximated by an integral. Instead, we approximate the binomial coefficients in the sum by its  maximum value, $i=(L-j)/2$, which is also in the central term of the sum. Expanding the  binomial coefficient using Stirling's approximation and performing the sum of the remaining terms, we obtain the approximation for large $j/L$ values,
\begin{equation}
    n_{L,j} = \frac{2^{L+j+2}}{6} \big[(L-j)^3 - (L-j) \big] \sqrt{\frac{2}{\pi (L+j+2)}} .
    \label{eqn:approx-2}
\end{equation}
In order to illustrate the quality of our approximations, we plot in Figure~\ref{fig:dist} the ratio between Eqn.~(\ref{eqn:dist}) and its approximations, Eqn.~(\ref{eqn:approx-1}) and Eqn.~(\ref{eqn:approx-2}).

\subsection{The mean number of reachable states}
\label{subsec:reachableliss}

Consider the {Preisach graph $(\vec{\alpha}, \vec{\omega})$} generated by the permutation $\vec{\rho}$. {As we have shown, this is the main hysteresis loop of the model and it is a maximal loop. Consequently, the reachable states of this loop are all $\rho$-stable, and we have denoted the set of these states  by $\mathcal{R}$.}   We would like to derive an expression for the number of reachable states this loop contains. From the discussion in Section \ref{sec:preigraph} it is clear that this is the number of leaves of the tree generated by the Preisach partition. 

Let {$\mathcal{N}(\vec{\rho}) = \vert \mathcal{R} \vert $} be the number of reachable states associated with the {Preisach graph}  generated by $\vec{\rho}$. Let $k$ be the position where $\rho_k = L$. Then with $\vec{\rho_-}$ and $\vec{\rho_+}$ as defined in \eqref{eqn:rhominusdef} and \eqref{eqn:rhoplusdef}, we have 
\begin{equation}
  \mathcal{N}(\vec{\rho}) = \mathcal{N}(\vec{\rho_-}) + \mathcal{N}(\vec{\rho_+}).
\label{eqn:prec}
\end{equation}
Recall our convention to consider the empty permutation as a Preisach system with $L = 0$, containing a single state, so that $\mathcal{N}(\emptyset) = 1$. Applying $\eqref{eqn:prec}$ recursively, the number of reachable states can be calculated for any permutation $\vec{\rho}$. Doing so, we are using in effect the partition tree generated by $\vec{\rho}$ as a substitution tree.

We will be interested in the mean number of reachable states $n_L$ of a Preisach system with $L$ hysterons. Since we are assuming that $\vec{\rho}$ is drawn uniformly from the set $\Pi_L$ of all permutations of $L$ elements, this is given by 
\begin{equation}
    n_L = \frac{1}{L!} \sum_{\vec{\rho} \in \Pi_L} \mathcal{N}(\vec{\rho}).
    \label{eqn:precmean}
\end{equation}
Using \eqref{eqn:prec} while conditioning on the position $k$ where $\rho_k = L$, and letting $k$ run from $1$ to $L$, the following recursion can be obtained
\begin{equation}
 n_L = 2 n_{L-1} - \frac{L -1}{L}\, n_{L -2}, 
\end{equation}
where $n_0 = 1$ and $n_1 = 2$. 

The numbers $n_L$ turn out to coincide with the mean number of increasing sub-sequences found in a permutation of the integers $\{1, 2, \ldots, L\}$ drawn uniformly at random \cite{AnalyticCombinatorics} (Example VIII.13 and VIII.43 p. 596-597). The asymptotic form of these numbers is given by \cite{lifschitz1981number,AnalyticCombinatorics}
\begin{equation}
    n_L = \frac{1}{2} \sqrt{\frac{1}{e \pi}}\, \frac{e^{2 L^{\frac{1}{2}} }}{L^{\frac{1}{4}}}. 
\end{equation}
Thus the leading order behavior of the mean number of reachable states in the {Preisach graph} grows with $L$ as a stretched-exponential. 
It turns out that a stronger connection between reachable states and increasing subsequences hold: 
\begin{itemize}
    \item [(P4)] For each permutation $\vec{\rho}$, the number of reachable states of the corresponding Preisach model is equal to the number of increasing sub-sequences contained in $\vec{\rho}$.
\end{itemize}

To see this, let $\Gamma(\vec{\rho})$ be the set of increasing sub-sequences contained in $\vec{\rho}$. This set can be partitioned into increasing subsequences containing the largest element $L$ and those that do not. If the largest element occurs at position $k$ of $\vec{\rho}$, so that $\rho_k = L$, then the set of increasing sub-sequences with and without $L$ are respectively equal to the number of increasing sub-sequences in the sub-permutations $\vec{\rho_-}$ and $\vec{\rho_+}$, with the latter given in terms of $\vec{\rho}$ by \eqref{eqn:rhominusdef} and \eqref{eqn:rhoplusdef}. Letting 
$ \vert \Gamma (\vec{\rho}) \vert$ denote the number of elements of $\Gamma(\vec{\rho})$ we therefore find that 
\begin{equation}
  \vert \Gamma (\vec{\rho}) \vert = \vert \Gamma (\vec{\rho_-}) \vert + \vert \Gamma (\vec{\rho_+}) \overset{}{},
\label{eqn:Grec}
\end{equation}
which is identical to \eqref{eqn:prec}. Moreover for the empty permutation we have $\vert \Gamma (\emptyset) \vert = 1$, so that the recursions \eqref{eqn:prec} and \eqref{eqn:Grec} are initialized with the same value and the equality $\mathcal{N}(\vec{\rho}) = \vert \Gamma (\vec{\rho}) \vert $ follows. The one-to-one correspondence between increasing sub-sequences and the reachable states of the {Preisach graph} generated by the same $\vec{\rho}$, is demonstrated in Fig.~\ref{fig:lRPM}(f), where below each state we have indicated the {corresponding subsequence}.

\section{Discussion}
\label{sec:discussion}

{The dynamics of the Preisach model is captured by its state transition graph, which describes the transitions between the configurations of its hysteretical elements as the driving field is changed just enough to trigger a state change. We have shown that the part of the transition graph that captures the transitions on  hysteresis loops and all their sub-loops, is actually 
not directly due to the particular values of the switching fields describing the switching behavior of the individual hysterons}, but a coarse-grained parameter, the permutation $\vec{\rho}$, 
which prescribes the sequence in which each of the hysterons return to its initial state relative to the order in which they were excited. These observations have led us to the notion 
of $\rho$-stability by identifying hysteron configurations that remain stable for all realizations of the switching fields that are compatible with $\vec{\rho}$. 
{We have shown that a state that is not $\rho$-stable cannot be part of any loop of the transition graph. We have called such states singleton states. While the presence or absence of singleton states in a transition graph depends -- besides $\vec{\rho}$ -- also on the particular values of the switching fields, the same loops and sub-loops will be present in the transition graph generated by any $\vec{\rho}$-compatible realization of the switching fields.}

An immediate consequence of $\rho$-stability is the robustness of {the loop and sub-loop} topology of the Preisach transition graph when interactions between hysterons are included, so that the switching fields of individual hysterons depend on the states of the other hysterons \cite{hovorka2005onset}. Such interactions occur between soft-spots in sheared amorphous solids and they are identifiable from numerical simulations \cite{mungan2019networks}. 
As long as these interactions are sufficiently weak, so that they do not alter the switching sequence $\vec{\rho}$, it then trivially follows from $\rho$-stability that the Preisach topology of the transition graph involving all states that are part of some hysteresis loop will prevail. 
The presence (or absence) of Preisach-like loop motifs in transition graphs arising from soft-spot systems, 
therefore provides a means by which one can infer the weakness (or strength) of the soft-spot interactions that realize these. Such Preisach-motifs have indeed been observed in transition graphs extracted from the simulation of sheared amorphous solids and under conditions where interactions between soft-spots are clearly present \cite{mungan2019networks}. 

The one-to-one correspondence between the states of the main hysteresis loop and the set of increasing subsequences contained in $\vec{\rho}$ is not a mere coincidence. It turns out that each increasing subsequence can be regarded as a field-history, providing  directions for how to reach a state via a sequence of field increases and decreases.
{The proof and further details on this correspondence have been given elsewhere \cite{ferrari2020liss}}. Here we note one interesting consequence: the length of the increasing subsequence is equal to the number of field-reversals and is therefore a physically meaningful quantity, characterizing the amount of memory that can be encoded via RPM  \cite{perkovic1997improved}. When the permutation $\vec{\rho}$ is drawn uniformly at random, it is known that the expected length of the longest increasing subsequence grows with $L$ to leading order as $2\sqrt{L}$ \cite{vershik1977asymptotics,logan1982variational,AnalyticCombinatorics}. It  can therefore be regarded as the typical amount of memory that can be encoded in a Preisach system via RPM \footnote{It is tempting to consider   the expected length of the longest increasing subsequence as an upper bound on the amount of memory that can be encoded in an interacting soft-spot system via RPM, since one would expect that the primary effect of adding interactions would to be a reduction of the number of stables states and the emergence of transitions between these involving more than one soft-spot change, {\em i.e.} avalanches. Both of these effects would tend to reduce the number of field-reversals of a field history leading to a reachable state. There are caveats though, one of them being that the leading order behavior of the length of the longest increasing subsequence was obtained under the assumption that the permutations are drawn at random and  uniformly. }.

\paragraph*{Acknowledgments}
The authors would like to thank Mahesh Bandi, Patrik Ferrari, Melih I{\c s}eri, Nathan Keim, Ido Regev, and Tom Witten for stimulating discussions. In addition, the authors also would like to acknowledge the comments and suggestions of the anonymous referees which have helped to improve the clarity of the presentation. MM was supported by the German Research Foundation (DFG) under DFG Projects No. 398962893 and 211504053, DFG Collaborative Research Center 1060 ``The Mathematics of Emergent Effects".

\bibliography{toy_model}%

\appendix

\section{$\rho$-stability of $2$-loops}
\label{app:2loop}

Suppose that $\vec{\sigma}$ satisfies the 2-loop condition \eqref{eqn:DUsigma}. From (P) it then follows 
that under $\Up$ and subsequent $\Dn$, a single hysteron $k$ first changes its state from $-1 \to 1$ and then back from $1 \to -1$. In order for this to occur it must be that 
\begin{equation}
    i^+[\vec{\sigma}] = i^-[\Up \vec{\sigma}] = k.
    \label{eqn:2loop}
\end{equation}
We first show that the choice of $\vec{\sigma}$ and $k$ satisfying the above condition depends entirely on $\rho$ and then show that such $\vec{\sigma}$ are stable for all choices of switching fields compatible with $\vec{\rho}$. 

Given $\vec{\rho}$, let us fix $k$ and ask for the configurations $\vec{\sigma}$ for which the condition \eqref{eqn:2loop} holds. From the ordering of the switching fields \eqref{eqn:Fplus_order} it follows that in order for $i^+[\vec{\sigma}] = k$, it must be that 
\begin{align}
    \sigma_k &= -1, \\
    \sigma_j &= +1,  \quad \mbox{for} \quad j < k.
\end{align}
This leaves the values of $\sigma_j$ for $j > k$ undetermined. Applying now 
$\Up$, it is clear that 
\begin{equation}
  (\Up \vec{\sigma})_i = + 1, \quad \mbox{for} \quad i \le k.
  \label{eqn:usigma}
\end{equation}
The subsequent $\Dn$ operation must change the value of hysteron $k$ back to $-1$. Let $r$ be the element of $\vec{\rho}$ for which $\rho_r = k$. The condition 
$i^-[\Up \vec{\sigma}] = k$ requires that for all $u \le r$, $\sigma_{\rho_u} = -1$. However by \eqref{eqn:usigma} the first $k$ hysterons must be in state $+1$. These two conditions can only be met, if
\begin{equation}
 \rho_u \ge k, \quad \mbox{for} \quad u \le r,  
\end{equation}
which, by uniqueness of the elements of $\vec{\rho}$, is equivalent to requiring $k = \rho_r$ to be such that 
\begin{equation}
     \rho_r = \min_{1 \le u \le r} \rho_u.
\end{equation}
In other words, $k$ must be a lower record of the sequence of elements of $\vec{\rho}$. Assuming such a choice of $k = \rho_r$, the configurations $\vec{\sigma}$ satisfying \eqref{eqn:DUsigma} must be of the form
\begin{equation}
    \sigma_i = \left \{ \begin{array}{cc}
    +1, & i < k, \\
    -1, & i \in \{ \rho_1, \rho_2, \ldots, \rho_r \}, \\
    *, & \mbox{otherwise},
    \end{array}
    \right.
    \label{eqn:2loop2app}
\end{equation}
with $k = \rho_r$ a lower record of $\vec{\rho}$. Note that  condition \eqref{eqn:2loop2app} depends only on $\vec{\rho}$ and that given the record-value $k$, there are $2^{L - k - r + 1}$ possible states $\vec{\sigma}$ that satisfy it.  
What remains to be  shown is that all such configurations $\vec{\sigma}$
are stable for any choice of $\vec{\rho}$-compatible switching fields,  
{\em i.e.} the stability condition \eqref{eqn:defStable} holds. 
First observe that $\vec{\sigma} = \vec{\alpha}$ is of the form  \eqref{eqn:2loop2} with $k = 1$ and thus $r$ is such that $\rho_r = 1$. The state $\vec{\alpha}$ is stable by assumption. With $\rho_r = k$ being a lower record of $\vec{\rho}$, consider a $\vec{\sigma} \ne \vec{\alpha}$ of the form given by \eqref{eqn:2loop2}. We have 
$F^+[\vec{\sigma}] = F^+_k$. Let us determine $F^-[\vec{\sigma}]$. Recall that $F^-[\vec{\sigma}]$ is given by 
\eqref{eqn:trapDefm}, which in turn depends on the set $I^-[\vec{\sigma}]$ of hysterons in state $+1$.
Since $\vec{\sigma} \ne \vec{\alpha}$, there is at least one site $j$ for which $\sigma_j = +1$. Let $j$ be any such hysteron.  Next, observe that the 
hysterons $\rho_1, \rho_2, \ldots, \rho_r = k$ are all in state $-1$, so that by \eqref{eqn:rhoorder} we must have
$F^-_j < F^-_k$ for all such $j$. Hence $F^-[\vec{\sigma}] < F^-_k < F^+_k = F^+[\vec{\sigma}]$ and we conclude that $\vec{\sigma}$ is stable.  

\section{Proof of the no-passing property for the Preisach Model}
\label{sec:NP}

In the context of the Preisach model and AQS dynamics, Middleton's no passing property reduces to finding a partial order $\preceq$ on the set of stable states $\mathcal{S}$ that is preserved by the dynamics. Let two initial configurations $\vec{\sigma}_1 \preceq \vec{\sigma}_2$ be given, respectively stable at initial forces $F_1(0) \le F_2(0)$. Assume that these configurations are subject to forces $F_1(t) \leq F_2(t)$. 
Denote the evolution of these states under their respective forces by $\vec{\sigma}_1(t)$ and $\vec{\sigma}_2(t)$. Middleton's no-passing property is 
the statement that $\vec{\sigma}_1(t) \preceq \vec{\sigma}_2(t)$ for all subsequent times $t$. 

Since we consider AQS dynamics, the evolution of $\vec{\sigma}_1(t)$ and $\vec{\sigma}_2(t)$ proceeds through a sequence of $\Up$ and $\Dn$-transitions under the influence of the driving forces. Since in the Preisach model hysterons change their state only from $-1$ to $1$ under $\Up$ (but not from $1$ to $-1$), and similarly from $1$ to $-1$ for $\Dn$-transitions, a natural partial order $\preceq$ on the set of 
stable states is the following:
\begin{equation}
I^+[\vec{\sigma}] \subset I^+[\vec{\sigma'}] \quad  \Leftrightarrow \quad    \vec{\sigma} \preceq \vec{\sigma'}.
\label{eqn:preisachPO}
\end{equation}
From the actions of $\Up$ and $\Dn$ as defined in Section \ref{sec:defs}, it is clear that for all configurations $\vec{\sigma}$, we have 
\begin{equation}
 \vec{\sigma} \preceq \Up \vec{\sigma},   
\end{equation}
with the equality holding only when $\vec{\sigma} = \vec{\omega}$ (in which case $ \Up \vec{\omega} = \vec{\omega}$). Likewise, we have 
\begin{equation}
\Dn \vec{\sigma} \preceq \vec{\sigma}, 
\end{equation}
again with the equality holding when $\vec{\sigma} = \vec{\alpha}$. 

We now turn to the proof of the no-passing property for the Preisach model. 
Assume that the no-passing property does not hold. We will show that this leads to a contradiction. Suppose that initially $\vec{\sigma}_1 \preceq \vec{\sigma}_2$ with the partial order defined by \eqref{eqn:preisachPO}. If the no-passing property fails, 
then there is a smallest time $t$ and a hysteron $j$ for which $\sigma_{2,j}(t) < \sigma_{1,j}(t)$ and hence $\sigma_{2,j}(t) = -1$ and $\sigma_{1,j}(t) = 1$.
This implies that prior to time $t$ the value of hysteron $j$ in both configurations must have been equal.
In the Preisach model the tipping fields $F^\pm_i$ of a hysteron $i$ depend only on the site, but not the full configuration. This implies that the 
tipping fields of $j$ are the same in both configurations. Suppose that prior to the flip hysteron $\sigma_j = -1$ in both configurations.   
Since $F_1(t) \leq F_2(t)$, it follows that if hysteron $j$ of the first configuration flipped from $-1$ to $1$, then certainly the same hysteron must have flipped also in the second configuration and thus a configuration with $\sigma_{2,j}(t) < \sigma_{1,j}(t)$ is impossible. 
The case when $\sigma_j = +1$ in both configurations prior to the flip leads to the same result. Thus the Preisach model has the no-passing 
property. Using the result of \cite{Sethna93}, the no-passing property then implies the return point memory 
property (RPM) .

\end{document}